\documentclass[pra,amssymb,amsmath,superscriptaddress,twocolumn,footinbib]{revtex4-2}
\usepackage{braket}
\usepackage{float}
\usepackage{tikz}
\usepackage{multirow}
\usepackage{booktabs,siunitx}
\usepackage[export]{adjustbox}
\usepackage{graphicx}
\usepackage{afterpage}
\usepackage[page,toc]{appendix}
\usepackage{amsthm}
\usepackage{dsfont}
\usepackage{hyperref}
\usepackage{chngcntr}
\hypersetup{
	colorlinks=true,
	linkcolor=blue,
	filecolor=magenta,      
	urlcolor=magenta,
	citecolor=teal
}
\usepackage{xcolor}
\usepackage[normalem]{ulem}
\usepackage{soul}

\newcommand{\boldm}{\boldsymbol{m}}
\newcommand{\boldM}{\boldsymbol{M}}
\newcommand{\outprod}[1]{\ket{#1}\!\!\bra{#1}}

\newcommand{\proj}[2]{\ket{#1}\!\!\bra{#2}}

\usepackage{afterpage}
\usepackage{scalerel}
\newcolumntype{P}[1]{>{\centering\arraybackslash}p{#1}}

\usepackage{dcolumn}
\usepackage{bm}
\usepackage{blkarray}
\usepackage{xparse}
\usepackage{mathtools}

\usepackage[shortlabels]{enumitem}

\usepackage{xcolor}

\begin{document}
	\preprint{APS/123-QED}
 \title{Entangling Quantum Memories at Channel Capacity}
	
	\author{Prajit Dhara}%
	\email[]{prajitd@arizona.edu}
	\affiliation{Wyant College of Optical Sciences, The University of Arizona, Tucson, AZ 85721}
	\affiliation{NSF-ERC Center for Quantum Networks, The University of Arizona, Tucson, AZ 85721}
	
	\author{Liang Jiang}
		\affiliation{NSF-ERC Center for Quantum Networks, The University of Arizona, Tucson, AZ 85721}
	\affiliation{Pritzker School of Molecular Engineering, University of Chicago, Chicago, IL 60637}
	
	\author{Saikat Guha}
	\email[]{saikat@umd.edu}
	\affiliation{Wyant College of Optical Sciences, The University of Arizona, Tucson, AZ 85721}
	\affiliation{NSF-ERC Center for Quantum Networks, The University of Arizona, Tucson, AZ 85721}
 \affiliation{Department of Electrical and Computer Engineering, University of Maryland, College Park, MD 20742}


\begin{abstract}
Entangling quantum memories, mediated by optical-frequency or microwave channels, at high rates and fidelities is key for linking qubits across short and long ranges. All well-known protocols encode up to one qubit per optical mode, hence entangling one pair of memory qubits per transmitted mode over the channel, with probability $\eta$, the channel's transmissivity. The rate is proportional to $\eta$ ideal Bell states (ebits) per mode. The quantum capacity, $C(\eta) = -\log_2(1-{\eta})$ ebits per mode, which $\approx 1.44\eta$ for high loss, i.e., $\eta \ll 1$, thereby making these schemes near rate-optimal. However, $C(\eta) \to \infty$ as $\eta \to 1$, making the known schemes highly rate-suboptimal for shorter ranges. We propose a cavity-assisted memory-photon interface that can be used to entangle matter memories with Gottesman-Kitaev-Preskill (GKP) photonic qudits, which along with dual-homodyne entanglement swaps that retain analog information, enables entangling memories at capacity-approaching rates at low loss. We benefit from loss resilience of GKP qudits, and their ability to encode multiple qubits in one mode. Our memory-photon interface further supports the preparation of needed ancilla GKP qudits. We expect our result to spur research in low-loss high-cooperativity cavity-coupled qubits with high-efficiency optical coupling, and demonstrations of high-rate short-range quantum links.
\end{abstract}

	\maketitle
	
	
    Entanglement is a key enabling resource for many tasks in quantum enhanced information processing. The task of generating entanglement at high rates and high fidelity between multiple parties is a core challenge, be it among multiple qubits in a quantum processor, between many nodes of a metropolitan-scale quantum network, or over long haul networks connecting quantum data centers relying on quantum repeaters, aerial nodes or satellite links. The fundamental unit of any such network is a single quantum link, connecting two parties over a lossy optical (or potentially microwave) frequency channel, often characterized by the channel's transmissivity $\eta$. 
    
    Most prevalent protocols for generating heralded entanglement among qubits in quantum memories connected by a lossy channel rely on quantum-optical modulation formats that encode up to one qubit per transmitted optical mode. This limits each transmitted mode to attempt to establish entanglement between at most a pair of distant qubits. If this attempt succeeds with probability proportional to $\eta$, e.g., when the dual-rail encoding is used which encodes one qubit using one photon spread across two modes, then the rate at which entanglement is generated is proportional to $\eta$ ideal Bell states (ebits) per mode. The maximum allowable entanglement rate, assisted with two-way authenticated classical communications, is the {\em quantum capacity} of the pure loss bosonic channel, given by: $C(\eta) = -\log(1-{\eta})$ ebits per mode. This scales roughly as $\approx 1.44\eta$ in the high loss, i.e., $\eta \ll 1$ regime, thereby making the known schemes near rate-optimal in the high-loss regime. However, $C(\eta) \to \infty$ as $\eta \to 1$. Therefore, for shorter ranges, viz., intra-processor and data-center quantum links, local-area quantum networks and potentially even long haul networks relying on ultra-low loss transmission assisted by vacuum beam guides~\cite{Huang2023-ta}, the known schemes are highly rate-suboptimal. 
    

    In this Article, we show that a cavity-assisted memory-photon interface we recently proposed~\cite{Dhara2024-ko} can be used to entangle $N$-qubit blocks of two remote memory registers, mediated by $d$-dimensional photonic (qudit) encodings at rates approaching capacity at low channel loss~\cite{Dhara2024-ko}. Our proposal uses individual devices that have been demonstrated in isolation, hence making it an attractive candidate for near-future realizations. We first generate -- at each of the two sites -- a maximally-entangled memory-photon entangled state of an $N$-qubit memory register and a single optical mode encoding $N = \log_2(d)$ qubits in the Gottesman-Kitaev-Preskill (GKP) photonic qu(d)it format~\cite{Gottesman2001-hd}. The optical qudits from each site undergo lossy propagation to a middle station where a dual-homodyne-based entanglement swap is performed, whose analog outcome is retained for further post-processing of the memory-memory remote entanglement created, at the error correction stage~\cite{Walshe2020-va,Fukui2021-ay}. If the channels have zero loss, and for an ideal cavity-assisted memory-photon interface, our protocol generates $N$ ebits per transmitted mode where $N$ can be arbitrarily high in principle, in congruence with the fact that $C(\eta) = \infty$ when $\eta = 1$. In the low-loss regime ($\eta$ approaching $1$), we show that the distillable entanglement rate achieved by our protocol approaches the quantum capacity, and outperforms known schemes by order of magnitude or more, even with finite-squeezed GKP qudits.
    

    Our scheme benefits from the resilience of GKP qudits to channel loss~\cite{Albert2018-qr,Noh2019-gd} which manifests as random Gaussian displacements in the phase space for GKP qudits~\cite{Gottesman2001-hd}. Furthermore, the GKP encoding can be naturally extended to accommodate multiple qubits in one mode, which is not possible with the more-common single-rail, dual-rail, cat-basis and single-photon multi-mode high-dimensional qubit encodings. It has also been shown that a similar memory-photon interface naturally supports the preparation of ancilla GKP qubits our scheme needs~\cite{Hastrup2022-wy,Weigand2018-lb}.
    Our results show that even with $5$-dB-squeezed GKP qubits, one could double intra-processor entanglement rates compared with the theoretical limits of known schemes utilizing single photon based photonic qubit encoding and linear optical entanglement swaps. Furthermore, by choosing an optimal encoding size $d$ commensurate to the channel loss, our protocol approaches capacity within a constant separation of $\sim\! 1$ ebit per channel use. We anticipate our result to lead to more sophisticated protocols and practical demonstrations of high-rate short-range quantum links. Extensions of our protocol are applicable to a variety of research questions including hybrid continuous-discrete variable quantum logic gates and resource efficient non Gaussian photonic state preparation. Furthermore, developing low-loss high-cooperativity cavity-coupled qubits with high-efficiency optical coupling is a major experimental target, and our scheme will further catalyze this technology development.
    
    Our article is structured as follows. We briefly introduce the GKP encoding and the underlying assumptions on our atomic quantum memories. We then describe the memory-photon interactions required to generate the spin-photon $d$-dimensional Bell pairs; implementing an effective CSUM gate. Finally, we evaluate the performance of the proposed quantum links while demonstrating their rate-optimality in the low-loss regime, and propose other potential applications of the CSUM protocol.

    \section*{Gottesman-Kitaev-Preskill Qudit Encoding}
    We consider a single bosonic mode with the creation ($\hat{a}$) and annihilation ($\hat{a}^\dagger$) operators defined in terms of the canonical position ($\hat{q}$) and momentum ($\hat{p}$) as $\hat{a}=(\hat{q}+i \hat{p})/\sqrt{2}; \hat{a}^\dagger=(\hat{q}-i \hat{p})/\sqrt{2}$ (i.e.\ $\hbar=1$ units). The logical \textit{square-lattice GKP} (sq-GKP) qudit subspace~\cite{Gottesman2001-hd,Noh2019-gd} is the eigenspace formed by the displacement operators:
    \begin{align}
        \begin{split}
            \hat{S}_q^{(d)}&=\exp{(-i\sqrt{2\pi d}\hat{p})}, \,{\text{and}}\\
            \hat{S}_p^{(d)}&=\exp(i\sqrt{2\pi d}\hat{q}).
        \end{split}
    \end{align}
    The logical basis states can be expressed in terms of position eigenstates as,
    \begin{align}
        \ket{\mu_q} \propto\sum_{n=-\infty}^{\infty} \ket{q=(dn+\mu_q)\sqrt{2\pi/d}},
        \label{eq:mu_q}
    \end{align}
    where $\mu_q=0,1,\ldots,d-1$. It is straightforward to note that a position quadrature displacement of $j\sqrt{2\pi/d}$, maps states from $\ket{\mu_q} \rightarrow \ket{\mu_q + j}$. Note that any logical state label $\mu_q$ is an integer in the residue set of the qudit dimension $d$; for the rest of article we will drop an implicit modulo in our description of the states. Another natural choice is the \textit{hexagonal-lattice GKP} (hex-GKP) qudit subspace, which is defined as the eigenspace of the displacement operators:
    \begin{align}
        \begin{split}
            \hat{S}_q^{(d)}&=\exp{[i\sqrt{2\pi d} \left({2}/{\sqrt{3}}\right)^{1/2} (\hat{q} +\hat{p}/2)]},\,{\text{and}}\\
            \hat{S}_p^{(d)}&=\exp{[-i\sqrt{2\pi d} \left({2}/{\sqrt{3}}\right)^{1/2} ( \sqrt{3}\hat{p}/2)]}.
        \end{split}
    \end{align}
    As with all GKP encodings~\cite{Gottesman2001-hd}, the ideal code states defined above are not physical owing to their non-normalizable definition, i.e., infinite superpositions of quadrature eigenstates. Physical approximations to the ideal states can be made by imposing finite peak widths and/or phase envelopes. In this Article, we will take finite-squeezed physical approximations as random-Gaussian displacements of the ideal logical states:
    \begin{align}
    	\begin{split}
    		\ket{\tilde{\mu}_q} \propto \int du \, dv \, \Gamma(u,v) \, \exp(-iu\hat{p})   \exp(iv\hat{q}) \ket{\mu_q},
    	\end{split}
     \label{eq:approx_GKP}
    \end{align}
    where $\Gamma(u,v)$ is a bi-variate Gaussian distribution,
    \begin{align}
    	\Gamma (u,v) = \frac{1}{2\pi {\sigma_u \sigma_v} }\exp\left[-\frac{u^2}{2\sigma_u^2} -\frac{v^2}{2\sigma_v^2} \right],
    \end{align}
    where typically $\sigma_u^2 =\sigma_v^2 =\sigma^2$.
    Although this definition is still not entirely practical (i.e., due to the infinite superposition in~\eqref{eq:mu_q}), the effect of channel loss and noise on GKP qudits can be effectively treated as evolution of the Gaussian-displacement variance $\sigma^2$, hence making it analytically tractable.
	
	\begin{figure}[h]
		\centering
		\includegraphics[width=0.8\linewidth]{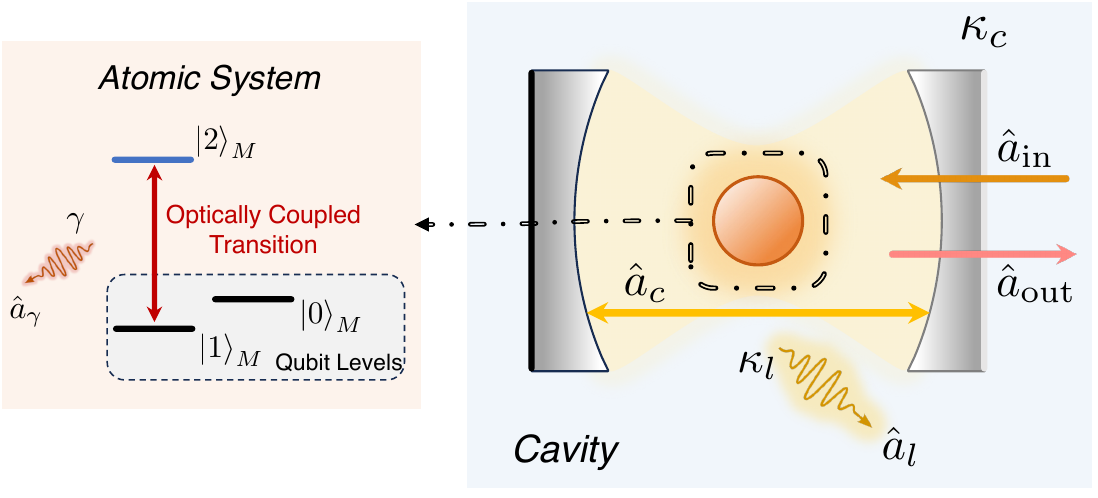}
		\caption{Overview of the cavity system coupled to the atomic 3-level system. The relevant modes, eigen levels and parameters are labeled. Detailed interaction models have been previously derived~\cite{Dhara2024-ko} (see \hyperref[sec:Methods]{Methods} for a summary). }
		\label{fig:cavity_atom_system}
	\end{figure}
    \section*{Cavity-Atom System}
 We consider $N$ three-level atomic systems (henceforth referred to as the {\em memories}) each with a single coupled interaction at an optical frequency (see Fig.~\ref{fig:cavity_atom_system}). Levels $\ket{0}_M$ and $\ket{1}_M$ form the logical qubit. The $\ket{1}_M$ level is optically coupled to a third level, $\ket{2}_M $. A variety of physical qubit platforms can support such a level structure, e.g., charged quantum dots~\cite{Najer2019-ed,Luo2019-ok}, trapped ions~\cite{Schupp2021-of}, cold atoms~\cite{Duan2004-qs}, and solid state vacancy centers~\cite{Bhaskar2020-kv,Stas2022-yw,Knaut2023-od}. Each atomic system is integrated in a single-mode cavity with one coupling port that couples the cavity mode ($\hat{a}_c$) with incoming ($\hat{a}_\mathrm{in}$) and outgoing ($\hat{a}_\mathrm{out}$) free-space optical modes at coupling rate $\kappa_c$. With the memory system initialized in its qubit subspace, the incoming optical field is reflected with a memory-state-dependent phase, effectively implementing an unitary transformation $CR_\pi$~\cite{Duan2004-qs,Hacker2019-ep}, where, 
	\begin{align}
		CR_\pi = \outprod{0}_M\otimes \hat{\mathbb{I}}_B +\outprod{1}_M \otimes \hat{R}_B{(\pi)},
	\end{align}
    with $\hat{R}_B{(\pi)}$ denoting a $\pi$-phase rotation on the reflected travelling wave bosonic (optical) mode. The controlled phase rotation gate's performance is degraded by the loss of photons from the cavity (at rate $\kappa_l$) and decay of the atomic state (from $\ket{2}_M \rightarrow\ket{1}_M$ at rate $\gamma_m$)~\cite{Duan2004-qs,Hacker2019-ep}. Practically, the performance of imperfect interactions is evaluated in terms of the system cooperativity, $C=4g^2/(\kappa\gamma_m) $ and the cavity efficiency, $\zeta=\kappa_c/(\kappa_c+\kappa_l)\equiv\kappa_c/\kappa$ where $g$ is the atom-cavity interaction strength. The detuning of the incoming photon (frequency $\omega$) from the cavity $(\omega_c)$ and the atomic transition $(\omega_a)$ are also relevant. The ideal operational regime is one where the excited state is virtually occupied, i.e., $\langle\outprod{2}_M-\outprod{1}_M\rangle\approx -1$, which is achieved for $|\omega-\omega_a|\gg g\sqrt{\langle \hat{a}^\dagger_c \hat{a}_c\rangle}$. It is important to note that for ${\langle \hat{a}^\dagger_c \hat{a}_c\rangle}\sim 1$ (approximately one photon interacts with the atom), one can derive exact dynamical equations for the field and atomic operators~\cite{Raymer2024-nz}. In a previous work~\cite{Dhara2024-ko}, we have derived the input-output relation of bosonic modes interacting with the joint atom-cavity system (see \hyperref[sec:Methods]{Methods} for an overview).

 \begin{figure*}
	\centering
	\includegraphics[width=0.85\textwidth]{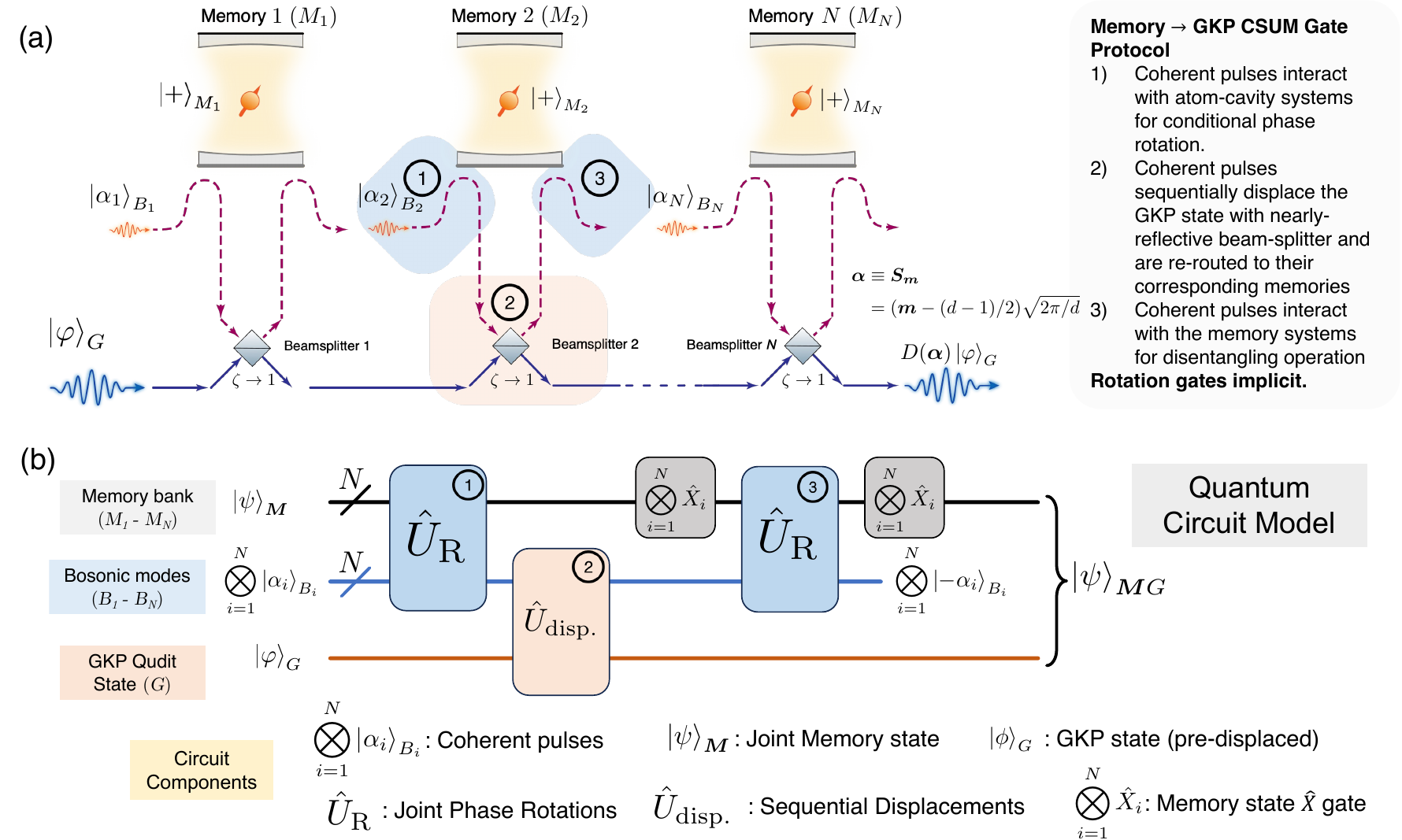}
	\caption{Overview of the interaction of the atomic system with the GKP qubit for the effective CSUM gate -- (a) Overview of the protocol implementing the CSUM gate, involving reflection of coherent pulses from a register of $N$-quantum memories integrated in cavities and memory-state dependent sequential displacement of pre-displaced GKP qudit states. (b) Circuit description of the protocol comprising of sequential controlled phase rotations $(\hat{U}_\mathrm{R})$ and sequential displacements $(\hat{U}_{\mathrm{disp.}})$.
	} 
	\label{fig:interaction_cav}
\end{figure*}

    \section*{Memory Register to GKP Qudit CSUM Gate Protocol}
    The controlled phase rotation of reflected photonic pulses described in the previous section is necessary to interface the memory register with a given GKP qudit state. For the description of our protocol, we will use the symbols $B_j, M_k$ and $G $ for the $j$-th interacting bosonic mode, $k$-th memory qubit and bosonic mode excited in the GKP qudit basis state respectively, where both $j,k\in\{1,2,\ldots,N\}$. In the interest of clarity, we describe the protocol to entangle sq-GKP qudits and memories in the main text, and prescribe the changes required to entangle hexagonal lattice GKP qudits afterwards.
    
    Without loss of generality, consider the case where the initial state of the GKP qudit is $\ket{0}_G$ (with $d=2^N$). The state is pre-displaced to the $\ket{\phi}_G= D((d-1)\sqrt{2\pi/ d}/2)\ket{0}_G$ state, { where $D(\alpha) = \exp(\alpha \hat{a}^\dagger_G -\alpha^* \hat{a}_G)$ is the displacement operator for the bosonic mode excited in the GKP qudit state}. Simultaneously, the memory register is initialized in the $ \ket{\psi}_{\boldsymbol{M}}=\bigotimes_{i=1}^N\ket{+}_{M_i}$ state. This can equivalently be expressed as
    \begin{align}
        \begin{split}
            \bigotimes_{i=1}^{N} \ket{+}_{M_i} =& \frac{1}{\sqrt{2^N}}\bigotimes_{i=1}^N(\ket{0}_{M_i}+\ket{1}_{M_i})\\=& \frac{1}{\sqrt{d}} \sum_{\boldm =0}^{d-1} \ket{\boldm}_{\boldM}.
        \label{eq:mem_joint_state}
        \end{split}
    \end{align}
    where, $\boldm =\{0,1,\ldots,d-1\}$ is the decimal representation of the $N$ bit binary string $m_N,m_{N-1},\ldots,m_1$ representing the joint states of the memory in the computational basis, ordered as $M_N,M_{N-1},\ldots,M_1$. 
    
    The CSUM gate protocol has three key steps which are as follows -- 
    \begin{enumerate}
    \item Reflect a pulse in the coherent-state $\ket{\alpha_k}_{B_k}$ of the $B_k$ mode off of the cavity-coupled memory $M_k$. The coherent amplitudes are specified by,
    \begin{align*}
        \alpha_k\sqrt{1-\zeta}=2^{N-k-1}\sqrt{2\pi/d}= d \times 2^{-k-1} \sqrt{2\pi/d}.
    \end{align*}
    Repeat this for all $k$.
    
    \item Sequentially apply beamplitter unitaries between the reflected modes and the mode with the GKP qudit state  nearly-reflective beamsplitters ($\zeta \rightarrow 1$) to perform sequential effective controlled displacements on the state.
    
    \item  Apply Pauli $\hat{X}$ gates on the memories prior to a second cavity reflection of the bosonic modes which nulls any memory-dependent phase imparted on the coherent state. 
    \item Repeat the Pauli $\hat{X}$ gates on the memories to generate the final state qudit entangled state between the memory register and the GKP state.
	\end{enumerate} 
     Considering the initial joint memory state $ \ket{\psi}_{\boldsymbol{M}}$ leads to the specific memory-GKP maximally entangled qudit state, i.e.,
     \begin{align}
         \ket{\Psi}_{\boldM G} =\frac{1}{\sqrt{d}} \sum_{m=0}^{d-1} \ket{\boldm}_{\boldM} \ket{m_q}_G,
         \label{eq:gkp_mem_max_ent}
     \end{align} 
     thereby implementing a memory register to GKP qudit CSUM gate
     
    The main feature of the ideal protocol (no imperfections in the cavity-photon interface) is that it applies a memory register state dependent displacement on the GKP qudit. For example, for a joint logical basis state $\ket{\boldm}_{\boldM}$, the interaction of the coherent pulses with the individual memories achieves the joint state transformation,
    \begin{align}
        \begin{split}
            & \ket{\boldm}_{\boldM} \otimes \left( \otimes_{k=1}^{N} \ket{\alpha_k}_{B_k}\right) \\
        &\;\rightarrow \ket{\boldm}_{\boldM} \otimes \left( \otimes_{k=1}^{N} \ket{(-1)^{m_k+1} \alpha_k}_{B_k}\right).
        \end{split}
    \end{align}
    Since the amplitudes are chosen appropriately to impart the desired displacements on the GKP state, the net displacement imparted by the conditionally rotated pulses $\otimes_{k=1}^{N} \ket{(-1)^{m_k+1} \alpha_k}_{B_k} $ (labelled by $\boldsymbol{S}$) is,
    \begin{align}
        \begin{split}
            \boldsymbol{S}_{\boldm} &=\sum_{k=1}^{N} (-1)^{m_k+1} \alpha_k \sqrt{1-\eta} \\
        &=  \sum_{k=1}^{N} (-1)^{m_k+1} \frac{d\sqrt{2\pi/d}}{2^{k+1}}\\
        &=  (\boldsymbol{m} - (d-1)/2)\times \sqrt{2\pi/d}.
        \end{split}
    \end{align}
    
    We note that the GKP state is pre-displaced -- hence, the net action of the protocol is to impart a memory state dependent displacement of $\boldsymbol{S}'_{\boldm} =\boldm \sqrt{2\pi/d}$, which yields the maximally entangled state as described in Eq.~\eqref{eq:gkp_mem_max_ent}. It is also important to note that by choosing amplitudes $\alpha_k e^{i\pi/2}$ and a commensurate state pre-displacement (i.e. along $p$ quadrature), this protocol can be modified to implement the $d$-dimensional analogue of the CZ gate. { For the hex-GKP encoding, the CSUM gate protocol is implemented by applying choosing appropriate pre-displacement and coherent amplitudes - readers may look at the 
    detailed proof for our specific choice of displacements in the \hyperref[sec:Methods]{Methods} (see Table~\ref{tab:disp_values}).}
    
	{ Any imperfection in the cavity photon interface limits the effectiveness of the proposed protocol. Primarily, the cavity-photon interface must impart the ideal memory-dependent phase while minimizing photon loss in the reflected bosonic pulse (either from imperfect cavity coupling or atomic state decay). Imperfect phases on the  reflected coherent pulses leads to 
    incorrect displacement of the GKP states, resulting in effective Pauli errors or states lying outside the GKP logical basis. Photon loss in the cavity-photon interface reduces the overall strength of the reflected pulses and results in an effective dephasing noise on the final state.} Considering these realistic effects, given the initial memory state initialization as in Eq.~\ref{eq:mem_joint_state}, the final state can most generally be described by,
	{\small
	\begin{align}
		\begin{split}
			\rho_{\boldM G}\approx \frac{1}{d} \sum_{\boldm, \boldm'=0}^{d-1} g_{\boldm, \boldm'}\proj{\boldm}{\boldm'}_{\boldM} D(\boldsymbol{\beta}_{\boldm'}) \rho_G D^\dagger(\boldsymbol{\beta}_{\boldm'}),
		\end{split}
		\label{eq:gkp_spin_ent}
	\end{align}} where corresponding to the memory basis state $\ket{\boldm}_{\boldM}$ the realistic displacement imparted is 
    \begin{align}
    \boldsymbol{\beta}_{\boldm}=\sum_{j=1}^{N} r^{(m_j)}(\omega) f(\omega) (-1)^{m_k+1} \alpha_{j}\sqrt{1-\zeta}.
    \end{align}
    Here, $r^{m_k}(\omega)$ describes the memory dependent phase imparted on the $k$-th coherent pulse with the spectral pulse shape $f(\omega)$. The coefficients $g_{\boldm, \boldm'}$ are given by,
    \begin{align}
        g_{\boldm, \boldm'} =\delta_{\boldm,\boldm'} + (1-\delta_{\boldm,\boldm'}) |\lambda_{\boldm,\boldm'}|^2 ,
    \end{align}
    with $|\lambda_{\boldm,\boldm'}|^2$  representing the effective dephasing factors due to photon loss and $\delta_{\boldm,\boldm'} $ is the Kronecker delta symbol. For $ C\gg1$ and $|\omega-\omega_c|\rightarrow0$, we may achieve the approximate relation $-r^{(0)*}(\omega)\approx r^{(1)*}(\omega)=1$ over the spectral bandwidth of the pulses (see \hyperref[sec:Methods]{Methods} for details). Any additional sources of loss can be incorporated by including sub-unity multiplicative factors for $|\lambda_{\boldm,\boldm'}|^2$. The efficacy of a non-ideal protocol implementation for $N=1; d=2$ (i.e. CNOT/CPHASE gate between a GKP qubit and a single memory) is discussed in detail in Ref.~\cite{Dhara2024-ko}.
	 
	For the required interaction to successfully apply the desired gate, we need to satisfy a few criteria for the interacting pulses. Assuming the cavity atom interface allows for $r^{(0)}(\omega)f(\omega )\rightarrow1$ and $r^{(1)}(\omega)f(\omega)\rightarrow-1 $, we would require the pulse with the largest coherent amplitude to satisfy the approximate displacement condition. By definition, the pulse in mode $B_1$ will be the largest; hence, we want $\pm \alpha_1 \sqrt{1-\eta}= \pm\sqrt{\pi d}/4 \Rightarrow \alpha_1 = \sqrt{\pi d/(1-\eta)}/4$. As the mean photon number content in the pulse is $|\alpha_1|^2= {\pi d}/{16(1-\eta)}$ we need the pulse length $\tau$ to satisfy $	\tau\gg{\pi d}/16\kappa(1-\eta) $ to ensure that the average photon number per interaction period is less than one for all pulses~\cite{Hacker2019-ep,Duan2004-qs}.

	
 

     \section*{Approaching Channel Capacity with Qudit Assisted Entanglement Swaps}

    \begin{figure}[ht]
		\centering
		\includegraphics[width=0.95\linewidth]{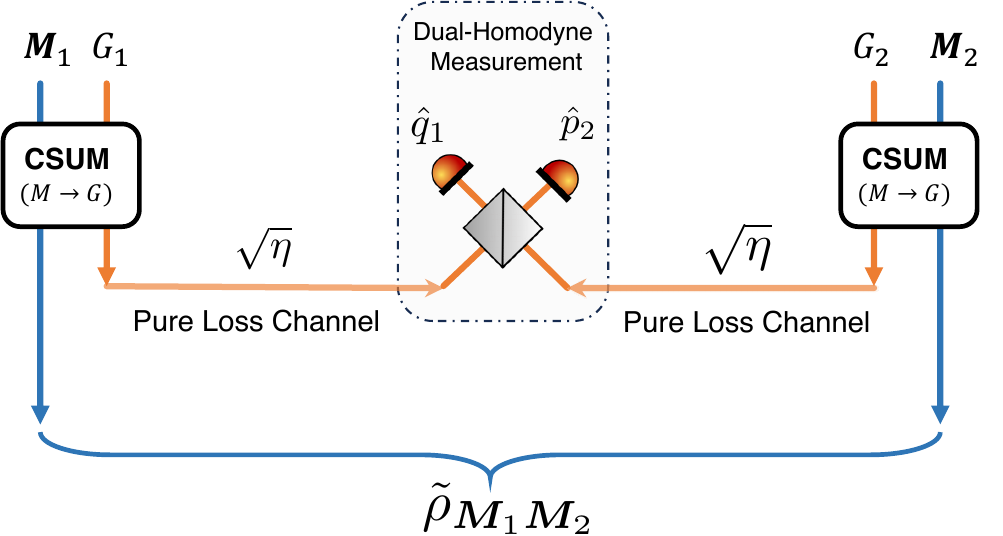}
		\caption{Layout of high-rate quantum links utilizing GKP qudit basis entanglement swap operation, where each party performs the CSUM gate on local memory registers $(\boldsymbol{M}_1,\boldsymbol{M}_2)$ and GKP qudit states $(G_1,G_2)$ before transmission to the midpoint of a link (of transmissivity $\eta$). The dual-homodyne measurements $\hat{q}_1, \hat{p}_2$ herald the generation of an entangled state of the memory registers $(\tilde{\rho}_{\boldM_1,\boldM_2})$. 
        }
		\label{fig:swap_layout}
    \end{figure}
     
     A variety of applications are enabled by the hybrid-entangled resource generation protocol discussed previously. The most straightforward application is bi-directional quantum state teleportation from either encoding/domain to the other. Our previous article Ref.~\cite{Dhara2024-ko} analyzes single qubit teleportation accounting for imperfect cavity-photon interfaces as well as realistic finite squeezed GKP qubit states. The resource state also enables the possibility of entangling two identical $N$-qubit memory registers by performing a projective GKP qudit basis Bell state measurement. 
     
     We assume the communicating parties have identical memory registers with the ability to perform the ideal CSUM gate interaction from the register to GKP qudit states. The parties first generate arbitrary memory-GKP maximally entangled states, and transmit the GKP qubit over identical bosonic pure-loss channels of transmissivities $\sqrt{\eta}$. The GKP qudit basis Bell state measurement~\cite{Walshe2020-va,Fukui2021-ay} is implemented on the states arriving at the midpoint, by interference on a 50:50 beamsplitter and dual-homodyne measurement (along orthogonal quadratures) on the pair of modes with real valued outcomes (Fig.~\ref{fig:swap_layout}). The memory registers are projected onto a maximally entangled state conditioned on these outcomes -- in principle, it is easy to see that we can choose $d$ optimally to enable high rate communications.
    
     The effect of loss {(with implicit pre-amplification of the state~\cite{Noh2019-gd} or post-processing~\cite{Fukui2021-ay} of measurement outcomes)} and finite-squeezing~\cite{Seshadreesan2022-bl} on the GKP qudit states can be effectively modeled by a Gaussian random displacement channel on the ideal qudit states~\cite{Fukui2021-ay,Schmidt2022-pj,Noh2019-gd}. Given an initial state described by the density operator $\rho_G$, the random displacement channel $\mathcal{N}_{B_2}[\sigma^2]$ modifies the state as
    \begin{align}
    	\mathcal{N}_{B_2}[\sigma^2](\rho_G)=\frac{1}{\pi\sigma^2} \int d^2\alpha\, e^{-|\alpha|^2/\sigma^2} D(\alpha) \rho D^\dagger(\alpha)
    \end{align}
    As the applied displacements are `continuous', the random displacement channel acting on a general GKP-encoded qudit state does not generally map to a higher dimensional analogue of a Pauli error channel. To simplify any treatment of the state, two strategies are commonly adopted by pre-existing literature -- the first being to consider the effect of noise and finite squeezing as being a coherent process, where the underlying peak variances of the GKP states are modified, and secondly, to `twirl' the random displacement channel to obtain a standard Pauli error channel. The former approach will yield an optimistic estimate (i.e.\ upper bound) of the protocol performance (because of the coherent treatment of errors). The latter approach neglects any residual coherences in the state and provides an achievable lower-bound -- we use this approach for our analysis.
    
      Although the dual homodyne measurement outcomes are a real-valued numbers, they can be post-processed for effective logical basis measurement outcomes on the GKP qudit basis, in addition to providing some analog information. Given real-value outcomes $x,y$ for homodyne measurements (along the $q$ and $p$ quadratures respectively), we can determine the measured logical state (for the corresponding measurement) by evaluating
     \begin{align}
     	\begin{split}
     	    x_L = \lfloor x/(\sqrt{2\pi/d})\rfloor \, (\text{mod } d); \\
     	y_L = \lfloor y/(\sqrt{2\pi/d})\rfloor \, (\text{mod } d),
     	\end{split}
     \end{align} 
     where $x_L, y_L\in\{0,1,\ldots,d-1\}$. The fractional remainders serve as analog information, which may be used for correction of the heralded state (by performing commensurate nulling displacements) and possible post-selection of the measurement outcome (which we do not perform). The fractional outcomes are evaluated as 
     \begin{align}
        \begin{split}
             x_f &= x/(\sqrt{2\pi/d}) - \lfloor x/(\sqrt{2\pi/d})\rfloor;\\
         y_f &= y/(\sqrt{2\pi/d}) - \lfloor y/(\sqrt{2\pi/d})\rfloor.
        \end{split}
     \end{align}
     For ideal GKP qudit states, the fractional outcome is always zero; however for realistic states, the fractional outcome is indicative of uncertainity in the peak centers (due to finite squeezing) or from peak position shifts due to the effective random displacement channel. 
     
     Let us assume that, as a virtue of possessing a perfect CSUM gate, ideal memories and ideal GKP qudits, both parties start with the ideal memory-GKP maximally entangled state,
     \begin{align}
         \ket{\Psi_{0,0}}_{\boldsymbol{M}_i,G_i} =\frac{1}{\sqrt{d}}\sum_{\boldm,m=0}^{d-1}\ket{\boldm}_{\boldM_i}\ket{m}_{G_i}; i=\{1,2\}. 
     \end{align}
    Lossless transmission i.e. $\eta=1$ yields a final state, $\ket{\Psi_{k',l'}}_{\boldM_1,\boldM_2}\equiv W_1^{(0,d-x_L)}W_2^{(d-y_L,0)} \ket{\Psi_{0,0}}_{\boldM_1,\boldM_2} $, a different qudit Bell state; since both $x_L$ and $y_L$ have $d$ possible values, the heralded state is one of $d^2$ qudit Bell states (see~\hyperref[sec:Methods]{Methods} for notation related to qudit gates).
     
     { Introducing any loss (even with ideal GKP states) yields a final state (say $\rho_{\boldM_1,\boldM_2}$) that is mixed.}  Assuming that the parties have access to finitely-squeezed GKP states of peak variance $\sigma_i^2$, transmission over a pure loss channel of transmissivity $\eta$ can be treated {as an effective transformation of the variance. We consider the variance transformation under two protocol choices - pre-amplification of the transmitted state (equivalent to applying an amplification channel of gain $G=1/\eta$ before transmission), or, classical post-processing of the dual-homodyne measurement outcomes by a classical computer, introduced as `classical computer assisted amplification' (\emph{CC-amplification}) in Ref.~\cite{Fukui2021-ay}. The transformations can be summarized as
     \begin{subequations}
         \begin{align}
         \text{Pre-amplification: }\sigma_i^2 \rightarrow \sigma_i^2+ (1-\eta) \\
         \text{CC-amplification: } \sigma_i^2 \rightarrow \sigma_i^2+ \frac{1-\eta}{2\eta}.
     \end{align}
     \end{subequations}
     It is straightforward to note that CC-amplification adds a smaller value of noise for all values of loss $\eta>0.5$. 
     } 
     
     Given a state variance of $\sigma^2$, the probability of making $k$-shift errors in the logical measurement for sq-GKP states (along either the Pauli $X$ or $Z$ basis) is given by~\cite{Noh2019-gd,Schmidt2024-br},
    {\small
    \begin{align}
         \begin{split}
             P^{(\mathrm{sq})}(X^{k},\sigma^2)&\equiv P^{(\mathrm{sq})}(Z^{k},\sigma^2) \\
         &= \sum_{j\in\mathbb{Z}} \frac{1}{2} \Biggl[\mathrm{erf}\left(\sqrt{\frac{2\pi}{d}} \frac{jd+k+1/2}{\sigma}\right) \\
         & \qquad \qquad \quad  - \mathrm{erf}\left(\sqrt{\frac{2\pi}{d}} \frac{jd+k-1/2}{\sigma}\right)\Biggr],
         \label{eq:shift_prob}
         \end{split}
     \end{align}
    }where $k\in\{0,1,2,\ldots,d-1\}$. $k=0$ denotes the probability of incurring no shifts; the summation over the index $j$ accounts for shifts of magnitude greater than one lattice spacing of $\sqrt{2\pi d}$. {Additionally, instead of exactly evaluating the effects of the random displacement channel, we consider a `twirled' noise model. This lets us approximate $\rho_{\boldM_1,\boldM_2}$ by the state,
    \begin{align}
    	\tilde{\rho}_{\boldsymbol{M}_1 \boldsymbol{M}_2} =  \sum_{k_1,k_2=0}^{d-1} P_{\mathrm{shift}}^{(k_1,k_2)}(\sigma^2)\,  \outprod{\Psi_{k',l'}}_{\boldsymbol{M}_1 \boldsymbol{M}_2}, 
    \end{align}
    where $P_{\mathrm{shift}}^{(k_1,k_2)}(\sigma^2)=P^{(\mathrm{sq})}(X^{k_1},\sigma^2)\cdot P^{(\mathrm{sq})}(Z^{k_2},\sigma^2)$.  Further analysis in this article is limited to $\tilde{\rho}_{\boldsymbol{M}_1 \boldsymbol{M}_2}$; we have elaborated on why this model of the output state serves as a mathematically simple and conservative estimate (in terms of the achievable entanglement generation rate)  for the final state in the~\hyperref[sec:Methods]{Methods}.} An achievable lower bound to the distillable entanglement for $\tilde{\rho}_{\boldsymbol{M}_1 \boldsymbol{M}_2}$ is given by the {\em hashing bound}~\cite{Devetak2005-hi},
    \begin{align}
    	I(\tilde{\rho}_{\boldsymbol{M}_1 \boldsymbol{M}_2}) = \!\log_2(d) + \!\!\!\! \sum_{k_1,k_2=0}^{d-1} P_{\mathrm{shift}}^{(k_1,k_2)}(\sigma^2)\log_2 P_{\mathrm{shift}}^{(k_1,k_2)}(\sigma^2).
    \end{align}

\begin{figure}[h]
		\centering
		\includegraphics[width=0.85\linewidth]{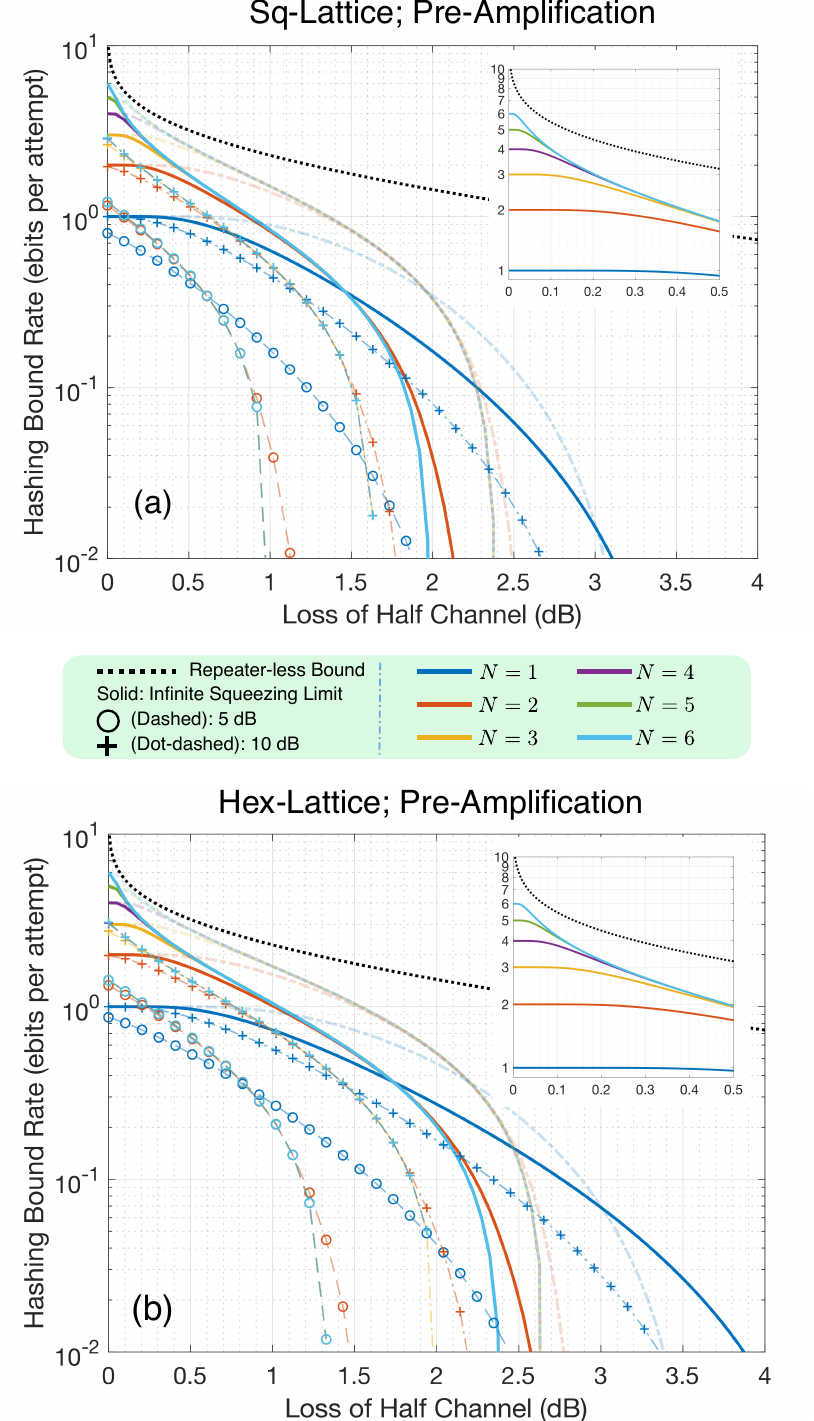}
		\caption{ Performance of the qudit-swap assisted link for $N$-qubit memory registers (varying colors) utilizing (a) square lattice  and (b) hexagonal lattice GKP qudit encodings $(d=2^N)$ assuming {pre-amplification of the transmitted qudits}. For both lattice choices, we analyze the performance of protocol for infinite squeezed states (solid), 5 dB (dashed; $\circ$ markers) and 10 dB (dot-dashed; $+$ markers) squeezed states with the channel capacity of $C(\eta)$.
        }
		\label{fig:teleport_plot_pre}
\end{figure}

\begin{figure}[h]
		\centering
		\includegraphics[width=0.85\linewidth]{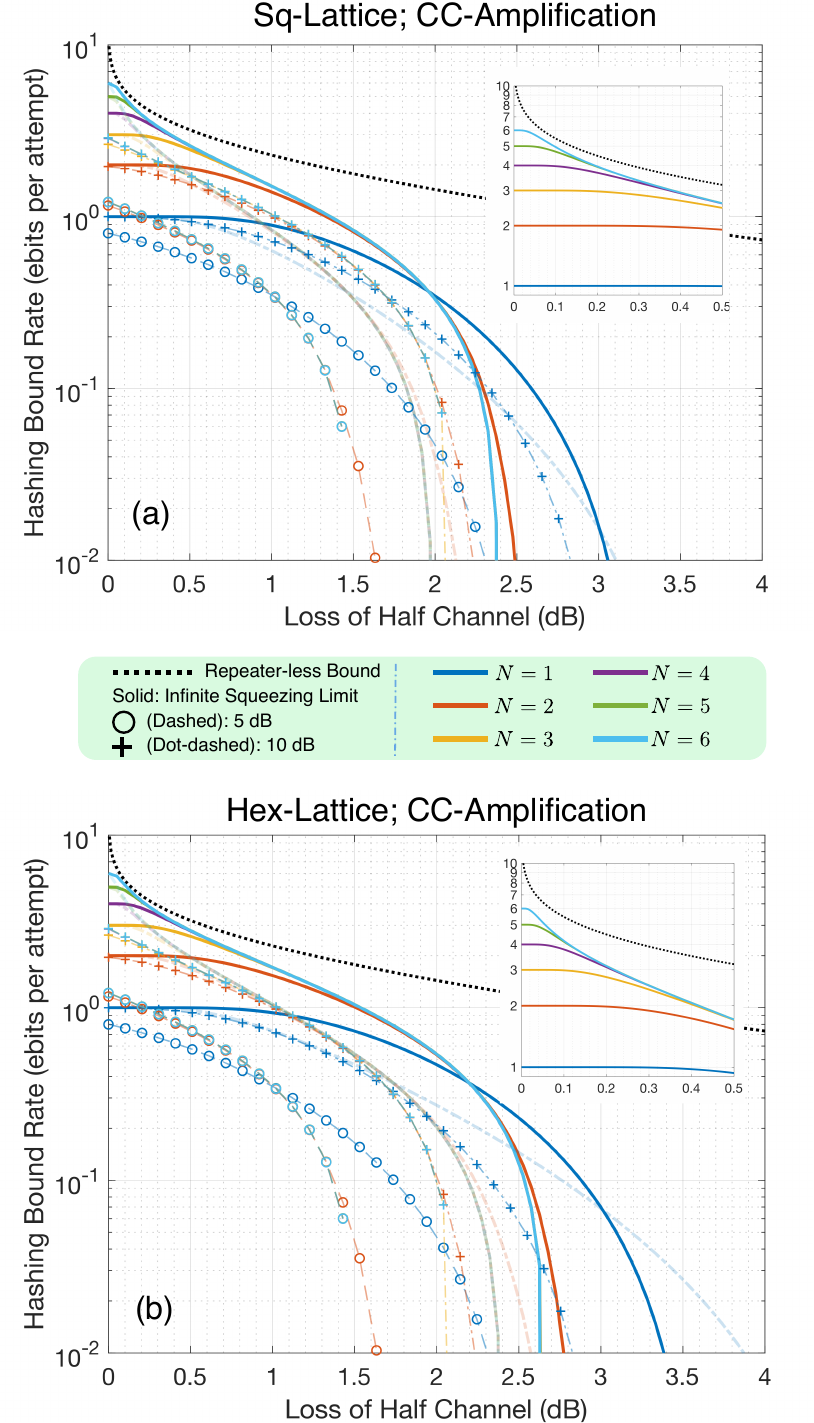}
		\caption{ Performance of the qudit-swap assisted link for $N$-qubit memory registers (varying colors) utilizing (a) square lattice  and (b) hexagonal lattice GKP qudit encodings $(d=2^N)$ assuming the {CC-amplification of dual-homodyne measurement outcomes}. For both lattice choices, we analyze the performance of protocol for infinite squeezed states (solid), 5 dB (dashed; $\circ$ markers) and 10 dB (dot-dashed; $+$ markers) squeezed states with the channel capacity of $C(\eta)$.
        }
		\label{fig:teleport_plot_cc}
\end{figure}

We plot the performance of the link for different values of the half-channel loss $(\sqrt{\eta})$ for various $N$-qubit memory registers (varying colors) utilizing GKP qudits of appropriate dimension $(d=2^N)$ in { Fig.~\ref{fig:teleport_plot_pre} (for pre-amplification of the transmitted state) and Fig.~\ref{fig:teleport_plot_cc} (for CC-amplification of the measurement results) for both the sq-GKP (a) and hex-GKP (b) encodings. The performance of infinite squeezed states (solid; $\sigma_i^2\rightarrow 0 $) is shown as a benchmark; finite squeezed states for 10 dB (dot-dashed; $+$ markers) and 5 dB (dashed; $\circ$ markers) of squeezing are shown for comparison. We compare these curves with the repeater-less channel capacity of $-\log_2(1-\sqrt{\eta})$ (black; dotted). In Fig.~\ref{fig:teleport_plot_pre} (\ref{fig:teleport_plot_cc}) we have included the performance of the CC-amplification (pre-amplification) protocol as semi-transparent color matched dot-dashed curves for various values $N$. }

For $\eta\rightarrow0$ dB, we note that $\mathcal{R}(\tilde{\rho}_{\boldsymbol{M}_1 \boldsymbol{M}_2}) \rightarrow N$. Thus by increasing the size of the memory registers, the rate scaling of $-\log_2(1-\sqrt{\eta})$ for $\eta\rightarrow1$ can be approached (with an offset of $\log_2 e$ similar to Ref.~\cite{Noh2019-gd}) by the proposed link. Furthermore the CC-amplification protocol (which adds lower variance noise), outperforms the pre-amplification for all values of half channel loss upto 3dB (as noted by the transformation rules earlier). Realistic implementations of the high-rate links will be affected by the CSUM gate implementation as well as the finite squeezing of the GKP states - we have analyzed some examples in the \hyperref[sec:Methods]{Methods}.

We also find that rate of entanglement with large dimensional encodings ($d\rightarrow\infty$) with infinitely squeezed GKP qudit states ($\sigma_i^2\rightarrow0$) approach the repeater-less bound $Q_2(\sqrt{\eta})= -\log_2(1-\sqrt{\eta})$ ebits per channel use for the entanglement swap in the low loss limit ($\eta \rightarrow 1$). Our detailed proof is included in the \hyperref[sec:Methods]{Methods} - we outline the highlights of the proof here --- In the limit of small state variance it is sufficient to lower bound the shift error probabilities $P(X^{k},\sigma^2) =P(Z^{k},\sigma^2)$ by dropping the summation over the lattice positions (i.e. $j$ index) in Eq.~\eqref{eq:shift_prob} - we may label this lower bound $\tilde{P}_{\mathrm{shift}}({k};\sigma^2)$. In the regime of interest, it is also sufficient to drop larger shifts by considering $\tilde{P}_{\mathrm{shift}}({k};\sigma^2)\approx 0$ for $|k|>1$, i.e.\ the probability of incurring larger (greater than one shift) logical errors is negligible. Carrying out a Taylor expansion of the probability of shift error around $d\rightarrow\infty$ (since we expect the system size requirements to increase) and $\eta\rightarrow1$, gives us the simplified expressions
\begin{align}
    \begin{split}
        &\tilde{P}^{(\mathrm{sq})}_{\mathrm{shift}}({0};\varepsilon)  = 1- \frac{\sqrt{2d\varepsilon} \exp\left(-\pi /(2d\varepsilon)\right)} {\pi}  \\
    &\tilde{P}^{(\mathrm{sq})}_{\mathrm{shift}}(\pm 1;\varepsilon) =  \frac{\sqrt{d\varepsilon} \exp\left(-\pi /(2d\varepsilon)\right)} {\sqrt{2}\pi}
    \end{split}
\end{align}
where $\varepsilon=1-\sqrt{\eta}$ {for the pre-amplification protocol}. We may make the ansatz that $2d\varepsilon=\xi$ where $\xi\in\mathbb{R}$ is an appropriately chosen constant that captures the required qudit dimensionality requirements. 

The hashing bound with all the approximations stated above (which we may label $I_{\mathrm{LB}} (\xi)$)  is lower than the hashing bound $I(\tilde{\rho}_{\boldsymbol{M}_1 \boldsymbol{M}_2})$. The matter of choosing an optimal $\xi$ to obtain the best rate scaling is equivalent to minimizing its difference from the channel capacity, i.e.\ we seek to minimize $Q_2(\sqrt{\eta})- I_{\mathrm{LB}} (\xi) $. This may be performed numerically; choosing $\xi_{\mathrm{opt}}\approx 1.642$ gives us a separation of $\sim1.06$ ebits per channel use between $I_{\mathrm{LB}} (\xi)$ and the repeaterless bound $Q_2(\sqrt{\eta})$. A similar ansatz can be made for the CC-amplification protocol where $\sqrt{\eta}=1-\varepsilon \Rightarrow (1-\sqrt{\eta })/(2\sqrt{\eta})\approx \varepsilon/2$. Then by choosing $d\varepsilon =\xi_{\mathrm{opt}}$ we obtain the same scaling as that of the pre-amplification protocol.

Further improvements in the entanglement swapping rate can be achieved by using the hex-GKP encoding. This is proven to be a more efficient lattice choice due to the efficient packing of lattice nodes in phase space. The logical shift error probability is modified in this case{
\begingroup\makeatletter\def\f@size{8}\check@mathfonts
\def\maketag@@@#1{\hbox{\m@th\large\normalfont#1}}%
\begin{align}
     \begin{split}
         &P^{(\mathrm{hex})}(X^{k},\sigma^2)\equiv P^{(\mathrm{hex})}(Z^{k},\sigma^2) \\
     &= \sum_{j\in\mathbb{Z}} \frac{1}{2} \left[ \mathrm{erf}\left(\sqrt{\frac{2\pi}{d}} \frac{jd+k+1/2}{\sigma(\sqrt{3}/2)^{1/2}}\right) - \mathrm{erf}\left(\sqrt{\frac{2\pi}{d}} \frac{jd+k-1/2}{\sigma(\sqrt{3}/2)^{1/2}}\right)\right],
     \label{eq:shift_prob2}
     \end{split}
\end{align}
\endgroup
}to account for the larger separation of $\sim \sqrt{4\pi/\sqrt{3}d}$ between the peaks of the logical states. We may then proceed similarly to the analysis carried out for the sq-GKP qubits - by ignoring greater than single lattice position shift errors, as well as making the appropriate Taylor series expansion, we obtain 
\begin{align}
    \begin{split}
        &\tilde{P}^{(\mathrm{hex})}_{\mathrm{shift}}({0};\varepsilon)  = 1- \frac{\sqrt{\sqrt{3}d\varepsilon} \exp\left(-\pi /(\sqrt{3}d\varepsilon)\right)} {\pi};  \\
     &\tilde{P}^{(\mathrm{hex})}_{\mathrm{shift}}(\pm 1;\varepsilon) =  \frac{\sqrt{\sqrt{3}d\varepsilon} \exp\left(-\pi /(\sqrt{3}d\varepsilon)\right)} {2\pi}.
    \end{split}
\end{align}
In this case, an ansatz of $\sqrt{3}d\varepsilon=\xi'$, allows us  to come up with a similar expression for $I_{\mathrm{LB}}^{(\mathrm{hex})}(\xi')$ and to optimize the rate lower bound separation from $Q_2(\sqrt{\eta})$. By choosing $\xi'_{\mathrm{opt}}\approx 1.422$ we obtain a separation of $\sim0.85$ ebits per channel -- this matches the expectation that $\xi_{\mathrm{opt}}/ \xi'_{\mathrm{opt}} = \sqrt{3}/2$, i.e. the qudit dimesionality vs.\ loss scaling is proportional to the improvement in the error probability between the lattice choices. Similarly for the CC-amplification, we may make the ansatz $\sqrt{3}d\varepsilon/2 = \xi'_{\mathrm{opt}}$.

\begin{figure}[h]
		\centering
		\includegraphics[width=0.95\linewidth]{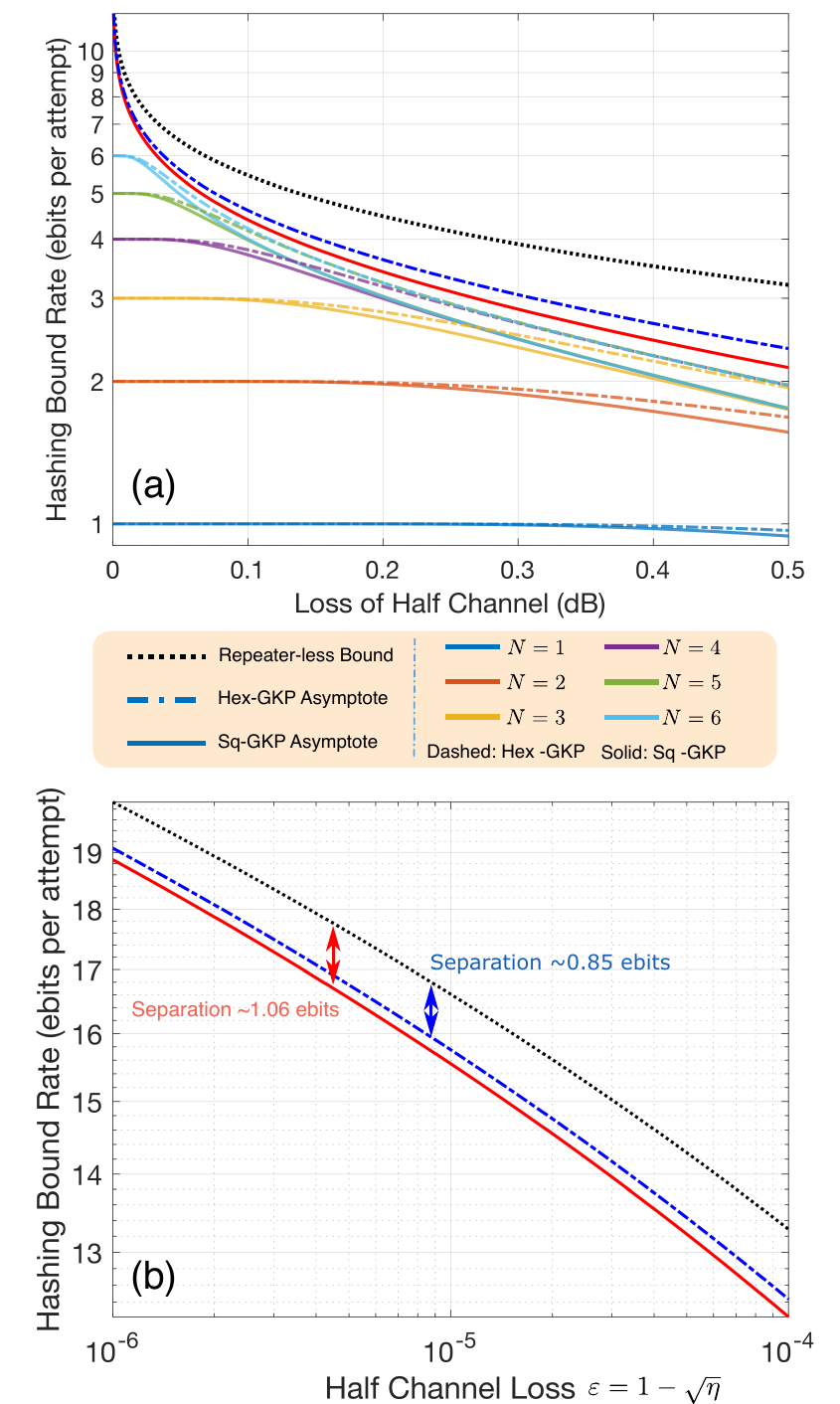}
		\caption{ Low-loss limit performance of the qudit-swap assisted link (a) Comparison of varying $N$ (varying colors) utilizing sq-GKP (solid lines)  and hex-GKP (dot-dashed) qudit encodings $(d=2^N)$ for the pre-amplification protocol. We compare the finite squeezing, finite encoding  size performance with calculated asymptotic rate scaling (red: sq-GKP; blue: hex-GKP) and the repeater-less channel capacity of $Q_2(\eta)=-\log_2(1-\sqrt{\eta})$ (black, dotted). (b) Comparison of the asymptotic expressions (for either protocol) and the repeater-less bound showing a finite separation in the extremely low-loss regime i.e.\ $\varepsilon\rightarrow 0; \eta\rightarrow 1$.
        } 
		\label{fig:low_loss_asymptote}
    \end{figure}

We contrast the performance of the link for different values of the half-channel loss $(\sqrt{\eta})$ for various $N$-qubit memory registers (varying colors) for both the sq-GKP (solid) and hex-GKP (dot-dashed) encodings, with their corresponding low-loss asymptotic curves in Fig.~\ref{fig:low_loss_asymptote}. In subfigure (b), we compare the asymptotic rate expressions with the repeater-less channel capacity and show the finite separation as predicted by our expressions.

    \section*{Outlook}
    We have presented a method for generating entanglement between a register of $N$-quantum memories and $d=2^N$ dimensional GKP qudit states, assisted by the interaction of bosonic modes with cavity QED systems. We demonstrated how the resource states generated by our protocol could be utilized for the implementation of high rate quantum links especially in the low-loss regime of operation, thereby highlighting a pathway to achieving channel capacity saturating quantum communication links.
    
    Furthermore, we show that the achievable entanglement generation rate scales as the repeaterless bound $C(\eta)$ for entanglement swapping in the low-loss regime with commensurately chosen $N$. State-of-the-art systems have demonstrated internal cooperativities of the order $\sim 100$. Improved cavity designs with larger cooperativities are being actively pursued in the various physical platforms that could support the necessary interactions. Coupled with efforts for large scale integration of physical quantum memory systems~\cite{Wan2020-op}, the protocols suggested here could be achievable in the near term. 
    
    The CSUM gate protocol proposed in this work is a key element for hybrid quantum logic between memory registers and GKP qudits -- the applications proposed in Ref.~\cite{Dhara2024-ko} can naturally be extended to high dimensional encodings. It is pertinent to note that protocols with higher-dimensional encodings are expected to be more sensitive to any imperfections in the cavity-photon interface. Further investigations of the underlying trade-offs and thresholds in the protocol performance are left for future works.

    \section*{Acknowledgements}
    The authors would like to thank Filip Rozpedek, Kaushik P. Seshadreesan, Christos N. Gagatsos, Aileen Zhai and Isack Padilla for fruitful discussions and comments on the manuscript. P.D. and S.G. acknowledge the Mega Qubit Router (MQR) project funded under federal support via a subcontract from the University of Arizona Applied Research Corporation (UA-ARC), for supporting this research. Additionally, all authors acknowledge the Engineering Research Center for Quantum Networks (CQN), awarded by the NSF and DoE under cooperative agreement number 1941583, for synergistic research support. L.J. additionally acknowledges support from the AFOSR MURI (FA9550-19-1-0399, FA9550-21-1-0209, FA9550-23-1-0338), DARPA (HR0011-24-9-0359, HR0011-24-9-0361), NSF (OMA-1936118, OMA-2137642, OSI-2326767, CCF-2312755), NTT Research, Packard Foundation (2020-71479), and the Marshall and Arlene Bennett Family Research Program.

    \section*{Conflicts of Interest}
    S.G. has outside interests in SensorQ Technologies Incorporated and Guha, LLC, and holds shares in Quantum Network Technologies (Qunett). These interests have been disclosed to UMD and reviewed in accordance with its conflict of interest (CoI) policies, with any resulting CoI to be managed accordingly.
    
	
	
	\onecolumngrid
 \appendix
 \section*{Methods}
 \label{sec:Methods}
    
    \subsection{Notation for Qudit Algebra and Gates}
    For a $d$-dimensional quantum system we may define a set of logical basis states as $\mathcal{L}_d=\{\ket{k}; k=0,1,\ldots,d-1\}$ such that $\braket{k|j}=\delta_{j,k}$. The corresponding Hadamard basis states is $\tilde{\mathcal{L}}_d=\{\ket{\tilde{l}}; \tilde{l}=0,1,\ldots,d-1\}$ such that $\braket{\tilde{m}|\tilde{n}}=\delta_{\tilde{m},\tilde{n}}$. Elements of the logical and Hadamard basis are related to one another by the Fourier relations,
    \begin{subequations}
            \begin{align}
            \ket{\tilde{l}}&=\frac{1}{\sqrt{d}} \sum_{k=0}^{d-1} e^{-i2\pi k \tilde{l}/d} \ket{k};\\
            \ket{k}&=\frac{1}{\sqrt{d}} \sum_{\tilde{l}=0}^{d-1} e^{i2\pi k \tilde{l}/d} \ket{\tilde{l}}.
        \end{align}
    \end{subequations}
    Similar to how the standard Pauli operators are defined for qubits, we can define generalized higher dimensional logical operators. The operators can be expressed either in terms of Gell-Mann matrices or using the Weyl operators~\cite{Bouda2001-fu} - we shall use the latter to denote the system operations. The standard \textit{Weyl operator} set for a $d$-dimensional qudit Hilbert space is defined as the set $\mathcal{W}=\{W^{(n,m)}| n,m =0,\ldots,d-1\}$, where,
        \begin{align}
            W^{(n,m)}_d =\sum_{k=0}^{d-1} e^{i2\pi k n/d} \proj{k}{k\oplus m}.
        \end{align}
    Here, $k\oplus m =k+m \,(\text{mod } d)$. It is important to note the following identities for the action of the Weyl operators
        \begin{enumerate}
            \item \textit{Identity operator}: $W_d^{(0,0)}=\sum_{k=0}^{d-1} \outprod{k}= \sum_{\tilde{l}=0}^{d-1} \outprod{\tilde{l}} \equiv \mathbb{I}_{d}$ i.e.\ $W_d^{(0,0)}$ is the $d$-dimensional identity operator.
            
            \item \textit{Computational basis traversal}: We have $ W^{(0,j)}_d \ket{{k}}= \ket{{k\ominus j}}$ where $k\ominus j =k-j \, (\text{mod } d)$. Hence, $ W^{(0,d-j)}_d \ket{{k}}= \ket{{k\ominus (d-j)}} =\ket{k\oplus j}$ where $k\oplus j =k+j \, (\text{mod } d)$. 
            
            \item \textit{Hadamard basis traversal}: We have $
            W^{(j,0)}_d \ket{\tilde{l}}= \ket{\tilde{l}\ominus \tilde{j}}$
            where $\tilde{l}\ominus \tilde{j}=l-j \, (\text{mod } d)$. Hence $W^{(d-j,0)}_d \ket{\tilde{l}}= \ket{\tilde{l}\ominus (d-\tilde{j})} =\ket{\tilde{l}\oplus \tilde{j}}$
            where $\tilde{l}\oplus \tilde{j}=l+j \, (\text{mod } d)$
        \end{enumerate}
    \textit{Maximally entangled states} of two $d$-dimensional qudit systems (labelled by their subscripts $1$ and $2$) are denoted by $\ket{\Psi_{k,l}}_{1,2}$, which are defined as,
			\begin{align}
				\ket{\Psi_{k,l}}_{1,2} = \frac{1}{\sqrt{d}} \sum_{k'=0}^{d-1} \exp(i2\pi k' l/d) \ket{k'}_{1}\otimes\ket{k'-k}_{2},
			\end{align}
			where $k,l=0,1,\ldots,d-1$. Correspondingly the \textit{positive operator valued measures (POVMs) for a qudit basis entanglement swap} are given by \begin{align}
       \Pi_\alpha =\{\outprod{\Psi_{k,l}}_{1,2}\,|\, k,l \in \{0,1,\ldots,d-1\}\}.
   \end{align}
    Given two maximally entangled states $\ket{\Psi_{k_1 ,l_1}}_{1,2}$ and $\ket{\Psi_{k_2 ,l_2}}_{3,4}$. Performing the projective measurement on systems $2,3$, with the outcome $\ket{\Psi_{r,s}}_{2,3}$ projects the systems $1,4$ in the joint state $\ket{\Psi_{k'l'}}_{1,4}$ where $k'=k_1+k_2-r\, (\text{mod }d)$ and $l'=l_1+l_2-s \, (\text{mod }d)$.

    \subsection{Detailed Model for Effective CSUM Gate Implemented Using Imperfect Cavity Photon Interactions }

    \subsubsection{Cavity-Photon Interaction Model}
    The effective CSUM gate protocol from a control system comprising of a register of $N$-quantum memories to a GKP qudit state of dimensionality $d=2^N$ relies on utilizing the effective control phase-rotation imparted on a coherent pulse by the interaction with a cavity-coupled atomic system. The controlled phase rotation applied to single photon states forms the basis of an atom-mediated polarization basis CPHASE gate proposed by Ref.\cite{Duan2004-qs}; extensions of this protocol have been suggested for the preparation of GKP grid states~\cite{Weigand2018-lb,Hastrup2022-wy}. Detailed derivation of the cavity-photon interaction (via the Heisenberg-Langevin formalism) can be found in Ref.~\cite{Dhara2024-ko}; here we state the main result and use it to derive our final state. Making the assumption that $\langle\hat{\sigma}_z\rangle\approx-1$ (i.e.\ the excited state $\ket{2}_M$ is only virtually populated for large detunings $\omega_a - \omega \gg g\sqrt{\braket{\hat{a}_c^\dagger \hat{a}_c} }$ ), one can show that the incoming and outgoing modes are coupled via the relation,
\begin{align}
	 \hat{a}_{\text {in}}(\omega)
	 &=r(\omega) \hat{a}_{\text {out}}(\omega)+l_C(\omega) \hat{a}_{l}(\omega)+ l_A(\omega) \hat{a}_{\gamma}(\omega),
\end{align} 
where, the mode transformation coefficients expressed in terms of the atom-cavity cooperativity $C=4g^2/(\kappa\gamma_m)$ and coupling efficiency $\xi=\kappa_c/(\kappa_c+\kappa_l)$ are given by 
\begin{subequations}
	\begin{align}
		r(\omega)&=1- \frac{2\zeta i}{\left(1-2i{\Delta_c}/{\kappa}+{C}/{(1-2i \Delta_{a}/\gamma_m)}\right)}; \\
		l_C(\omega)&=-\frac{2\sqrt{\zeta\,(1-\zeta)}}{\left(1-2i {\Delta_c}/{\kappa}+{C}/{(1-2i \Delta_{a}/\gamma_m)}\right)}; \\
		l_A(\omega)&=\frac{{-2i\sqrt{\zeta C}/(1-2i\Delta_a/\gamma_m)}}{\left(1-2i {\Delta_c}/{\kappa}+{C}/{(1-2i \Delta_{a}/\gamma_m)}\right)}.
	\end{align}
\end{subequations}
In any interaction, modes represented by $\hat{a}_\gamma $ and $\hat{a}_l$ will be lost since they represent scattered field operators - hence it is appropriate to call $r(\omega)$ the reflectivity, $l_C(\omega)$ the cavity loss coefficient and $l_A(\omega)$ the atomic loss coefficient. An improved approach with similar conclusions is shown in a recent work~\cite{Raymer2024-nz}.

\subsubsection{Pulse Interaction}
Here-on we use the subscripts $B_j, M_k$ and $G $ for the $j$-th bosonic mode, $k$-th memory qubit and GKP qudit (or rather the bosonic mode that the qudit occupies) respectively. We begin by considering the interaction of a coherent pulse $\ket{\alpha_j}_{B_j}$ excited in the spectral mode $f_j(\omega)$ -- additionally for ease of  analysis, we introduce the following equivalent notation for the coherent pulse to track the spectral response of the atomic system,
\begin{align}
	\ket{\alpha}_{B_j}
 &\equiv \bigotimes_{\omega} \ket{\alpha_j f_j(\omega) d\omega}_{B_j}.
\end{align}
The interaction of the pulse with an arbitrary memory state $\ket{\psi_j}_{M_j}=c_{j,0}\ket{0}_{M,j} +c_{1,j} \ket{1}_{M,j}$ where $c_{j,0},c_{j,1}\in \mathbb{C}$ and $|c_{j,0}|^2+|c_{j,1}|^2=1$ yields the output state
\begin{align}
    \ket{\phi}_{M_j,B_j}= \sum_{s=\{0,1\}} c_{j,s}\ket{s}_{M,j} \bigotimes_{\omega} \ket{r^{(s)} (\omega) \alpha_j f_j(\omega) d\omega}_{B_j} \ket{l_C^{(s)} (\omega) \alpha_j f_j(\omega) d\omega}_{L_{C,j}} \ket{l_A^{(s)} (\omega) \alpha_j f_j(\omega) d\omega}_{L_{A,j}},
\end{align}
where $s$ denotes the state of the memory and modes $L_{C,j}, L_{A,j}$ are the cavity and atomic loss modes. Let us now write down the final state considering a collective state $\ket{\boldm}_{\boldM}$ of the memory register. $\ket{\boldm}_{\boldM}$ are the joint basis states for the memory register,  where $\boldm$ is the decimal representation of the $N$ bit string for the memories ordered as $M_N,M_{N-1},\ldots,M_1$, where the state of $M_N$ is the least significant bit of $\boldm$ . The interaction of the cavity pulses with the memories yields the final state
\begin{align}
    \ket{\phi_{\boldm}}_{\boldM,\boldsymbol{B}}= \ket{\boldm}_{\boldM} \bigotimes_{j=1}^{N} \bigotimes_{\omega} \ket{r^{(m_j)} (\omega) \alpha_j f_j(\omega) d\omega}_{B_j} \ket{l_C^{(m_j)} (\omega) \alpha_j f_j(\omega) d\omega}_{L_{C,j}} \ket{l_A^{(m_j)} (\omega) \alpha_j f_j(\omega) d\omega}_{L_{A,j}}
\end{align}
Thus starting with initial memory state of $\ket{\Psi}_{\boldM}=\sum_{\boldm} c_{\boldm} \ket{\boldm}_{\boldM}  $ yields the final state $\ket{\Phi}_{\boldM,\boldsymbol{B}}=\sum_{\boldm} c_{\boldm} \ket{\phi_{\boldm}}_{\boldM,\boldsymbol{B}}$, where upon tracing out the cavity and atomic loss modes we obtain the state,
\begin{align}
    \rho_{\boldM,\boldsymbol{B}} =\frac{1}{d} \sum^{2^N}_{\boldm,\boldm'=1} g_{\boldm,\boldm'} \;\proj{\boldm}{\boldm'}_{\boldM}\otimes \proj{\beta_{\boldm}}{\beta_{\boldm'}}_{\boldsymbol{B}},
    \label{eq:cav_pulse_1int}
\end{align}
where $\beta_{\boldm}$ represents the joint reflected pulse amplitude, $g_{\boldm,\boldm'}$  the state density matrix element for position ${\boldm,\boldm'}$ and $\lambda_{\boldm,\boldm'}$ the effective dephasing terms. These can be calculated by evaluating 
\begin{subequations}
    \begin{align}
    \beta_{\boldm} & = \bigotimes_{j=1}^{N} \bigotimes_{\omega} \ket{r^{(m_j)} (\omega) \alpha_j f_j(\omega) d\omega}_{B_j};\\
    g_{\boldm,\boldm'} & =c_{\boldm} c^*_{\boldm'} \left( \delta_{\boldm',\boldm'} + (1-\delta_{\boldm',\boldm'}) \lambda_{\boldm',\boldm'} \right);\\
    \lambda_{\boldm',\boldm'} & = \bigotimes_{j=1}^{N} \bigotimes_{\omega} \braket{l_C^{(m_j)} (\omega) \alpha_j f_j(\omega) d\omega|l_C^{(m'_j)} (\omega) \alpha_j f_j(\omega) d\omega}_{L_{C,j}} \times \braket{l_A^{(m_j)} (\omega) \alpha_j f_j(\omega) d\omega|l_A^{(m'_j)} (\omega) \alpha_j f_j(\omega) d\omega}_{L_{A,j}}.
\end{align}
\end{subequations}
Thus by extension the state of the memory-pulse state after the second interaction (after the sequential displacement circuit and the memory $\hat{X}$ gates) can be equivalently expressed by Eq.~\eqref{eq:cav_pulse_1int}, where the $ g_{\boldm,\boldm'}$ are modified as 
\begin{align}
    g_{\boldm,\boldm'} & =c_{\boldm} c^*_{\boldm'} \left( \delta_{\boldm',\boldm'} + (1-\delta_{\boldm',\boldm'}) |\lambda_{\boldm',\boldm'}|^2 \right).
\end{align}

\subsubsection{Memory State Dependent Displacement of GKP Qudit State}
Given a joint memory register state denoted by $\ket{\boldm}_{\boldM}$, the sequence of $N$-coherent pulses $\ket{\alpha_1} - \ket{\alpha_N}$ in modes $B_1-B_N$ are modified as $\ket{\alpha_k} \rightarrow \ket{(-1)^{m_k} }$, where $m_k$ is the $k$-th least significant bit in the binary represenation of $\boldm$. Assuming we choose $\alpha_k$ such that $\alpha_k\sqrt{1-\zeta}=2^{N-k-1}\sqrt{2\pi/d}= d \times 2^{-k-1} \sqrt{2\pi/d}$ (for the effective memory to sq-GKP CSUM gate), the action of the sequential displacement operations using the nearly transmissive beamsplitters ($\eta\rightarrow1$) is given by an effective net displacement operator on the GKP qudit state by $\boldsymbol{\mathcal{S}}_{\boldm}$ given by
\begin{subequations}
    \begin{align}
    \boldsymbol{S}_{\boldm} = & \sum_{k=1}^{N} (-1)^{m_k+1} \alpha_k \sqrt{1-\eta} 
    = \sum_{k=1}^{N} (-1)^{m_k+1} d \times 2^{-k-1} \sqrt{2\pi/d} \\
    = &  \frac{d \sqrt{2\pi/d}}{ 2} \, \sum_{k=1}^{N} (-1)^{m_k+1} \times 2^{-k}  \\ 
    =& \frac{\sqrt{2\pi/d}}{ 2} \, \sum_{k=1}^{N} (m_k -\bar{m}_k) \times 2^{N-k} 
    =  \frac{\sqrt{2\pi/d}}{ 2} (\boldm - \bar{\boldm}) \label{eq:disp_deriv1}\\
    = & \frac{\sqrt{2\pi/d}}{ 2} (2 \boldm - (d-1)) = (\boldsymbol{m} - (d-1)/2)\times \sqrt{2\pi/d} .\label{eq:disp_deriv2}
\end{align}
\end{subequations}
Here, we use the binary representation of $\boldm = [m_1 m_{2} \ldots m_N]_2 = \sum_{k=1}^{N} m_k \times 2^{N-k} $. Additionally in Eq.~\eqref{eq:disp_deriv1}, we first use the equivalence between the functions $g(m_k)= (m_k -\bar{m}_k)$ and $f(m_k)= (-1)^{m_k+1}$ for $m_k=\{0,1\}$, where $\bar{m}_k$ is the complement of ${m}_k$. We use $\bar{\boldm}$ to represent the decimal representation of an integer obtained by bitwise complement of the bits of $\boldm$, i.e., $\bar{\boldm} = [\bar{m}_1 \bar{m}_{2} \ldots \bar{m}_N]_2$. In Eq.~\eqref{eq:disp_deriv2}, we use the identity $\bar{\boldm} = (2^N-1) - \boldm =(d-1)-\boldm$ to obtain the final expression. We list the necessary coherent amplitudes, and pre-displacements for the various gates that can be implemented by the proposed interactions in Table~\ref{tab:disp_values}.

\begin{table}[h]
    \centering
    \begin{tabular}{>{\centering}p{2cm}>{\centering}p{2cm}>{\centering}p{4cm}>{}p{3cm}}
        \toprule
         \textbf{Encoding} & $\boldM \rightarrow G$ \textbf{Gate}  &  \textbf{Coherent Amplitudes} $\alpha_k\sqrt{1-\zeta}$ & \textbf{Pre-Displacement Amplitude} \\
         \midrule
         \multirow{ 2}{*}{sq-GKP} & CSUM & $ \frac{d}{2^{k+1}} \sqrt{\frac{2\pi}{d}}$ & $ \frac{d-1}{2} \sqrt{\frac{2\pi}{d}}$ \\
         & CPHASE & $ \frac{id}{2^{k+1}} \sqrt{\frac{2\pi}{d}}$ & $ \frac{i(d-1)}{2}  \sqrt{\frac{2\pi}{d}}$ \\
         \midrule
         \multirow{ 2}{*}{hex-GKP}& CSUM & $ \frac{d}{2^{k+1}} \sqrt{\frac{2\pi}{\sqrt{3}d}} \left( \frac{\sqrt{3}-i}{2}\right)$ & $ \frac{d-1}{2} \sqrt{\frac{2\pi}{\sqrt{3}d}} \left( \frac{\sqrt{3}-i}{2}\right)$ \\
         & CPHASE & $ \frac{id}{2^{k+1}} \sqrt{\frac{2\pi}{\sqrt{3}d}} $ & $ \frac{i(d-1)}{2} \sqrt{\frac{2\pi}{\sqrt{3}d}} $ \\
         \bottomrule
    \end{tabular}
    \caption{Required coherent amplitudes for reflected pulses and pre-displacement to implement various memory register to GKP ($\boldM \rightarrow G$) qudit gates.}
    \label{tab:disp_values}
\end{table}

\subsection{Bounding the Quality of Distributed Entanglement}
{
A faithful analysis of the proposed quantum link would require a complete evolution of the transmitted GKP qudit states through the pure loss channel (with possible pre-amplification).
However, given the non-Gaussian state description of the underlying states, this process is complicated. Analyses by prior studies~\cite{Rozpedek2023-vi,Fukui2021-ay} consider a `twirled' error approach - we present an argument why this approach presents a conservative estimate to the final state quality.

We implicitly assume that the pure loss channel is augmented by amplification (either pre-amplification of state of post-amplification of the measurement outcomes), thereby effectively converting it into a random displacement channel, $\mathcal{N}_{B_2}[\sigma^2]$. Given an initial pure approximate GKP qudit state $\ket{\chi}_G$ , the channel output is described by
    \begin{align}
    	\mathcal{N}_{B_2}[\sigma^2](\rho_G)=\frac{1}{\pi\sigma^2} \int d^2\alpha\, e^{-|\alpha|^2/\sigma^2} D(\alpha) \outprod{\chi}_G D^\dagger(\alpha).
    \end{align}
Rather than evaluating the continuous integral over phase space, a few simplified approaches are possible - 
\begin{enumerate}
	\item We may treat the effect of random displacement as an effective modifier to the individual peak variance for the GKP state while preserving initial state purity. Given the initial state $\ket{\chi}_G$ expressed in the approximate GKP basis, Eq.~\eqref{eq:approx_GKP} with some initial variance $\sigma^2_i$, the final state is a pure state $\ket{\chi'}_G$ with a modified variance $\sigma_i^2 + g(\sigma^2)$, where the modifier $g(\sigma^2)$ is a function of the random noise parameter. 
	
	\item We may `twirl' the channel and approximate the continuous random displacements as discretized Pauli errors with a weighted error probability. This is described by 
    \begin{align}
        \frac{1}{\pi\sigma^2} \int d^2\alpha\, e^{-|\alpha|^2/\sigma^2} D(\alpha) \outprod{\chi}_G D^\dagger(\alpha)  \rightarrow & \sum_{k_1,k_2} \int_{\mathbb{L}(k_1,k_2)} d^2\alpha\, e^{-|\alpha|^2/\sigma^2} D(\alpha) \outprod{\chi}_G D^\dagger(\alpha) \nonumber \\ &\approx \sum_{k_1,k_2} P(X^{k_1}) P(Z^{k_2}) \hat{X}^{k_1} \hat{Z}^{k_2} \outprod{\chi}_G \hat{Z}^{k_2} \hat{X}^{k_1},
        \label{eq:twirled_model}
    \end{align}
    where $\mathbb{L}(k_1,k_2)$ is the lattice tile corresponding to $k_1$ shifts along the logical Pauli $\hat{X}$ and $k_2$ shifts along the logical Pauli $\hat{Z}$ directions for the qudit on phase space. The probabilities $P(X^{k_1}) $ and $P(Z^{k_2})$ correspond to the integrated probability over this lattice tile, while $\hat{X}^{k_1} = W_d^{(0,k_1)}$ and $\hat{Z}^{k_2} = W_d^{(k_2,0)}$ are the qudit basis traversal operators. 
 
\end{enumerate}
Assuming both parties start with the ideal memory-GKP maximally entangled state,
     \begin{align}
         \ket{\Psi_{0,0}}_{\boldsymbol{M}_i,G_i} =\frac{1}{\sqrt{d}}\sum_{\boldm=m=0}^{d-1}\ket{\boldm}_{\boldM_i}\ket{m}_{G_i}. 
         \label{eq:ideal_init_state}
     \end{align}
Assuming ideal GKP qudits, and no loss in transmission, the dual homodyne measurement based Bell state measurement, heralds a pure state $\ket{\Psi_{k,l}}_{\boldsymbol{M}_1,\boldM_2}$. If the GKP qudits are transmitted over a link of transmissivity $\eta <1$, thereby undergoing the aforementioned random displacement channel, the final state $\rho_{\boldM_1,\boldM_2}$ will be a mixed state. 

We now consider the approximate models to random displacement. Assuming the pure state evolution noise model, we label the heralded pure state $\ket{\Psi'}_{\boldsymbol{M}_1,\boldM_2}$. However with the twirled noise model, the final state is mixed and labelled  $\tilde{\rho}_{\boldM_1,\boldM_2}$. The pure state evolution model treats the effect of loss coherently, hence, the entanglement distribution rate estimated by this approach would overestimate the actual performance. One the other hand, the twirled error model ignores residual coherences between the `shifted' states - we ignore terms of the form $\hat{X}^{k_1} \hat{Z}^{k_2} \outprod{\chi}_G  \hat{Z}^{k'_2} \hat{X}^{k'_1}$, where $k_1\neq k_1'; k_2\neq k_2'$ in Eq.~\eqref{eq:twirled_model}. Hence we expect the following inequality,
\begin{align}
    I(\outprod{\Psi'}_{\boldM_1,\boldM_2}) \geq I(\rho_{\boldM_1,\boldM_2}) \geq I(\tilde{\rho}_{\boldM_1,\boldM_2})
\end{align}
where $I(\rho_{AB}) = \max[S(\rho_A)-S(\rho_{AB}),S(\rho_B)-S(\rho_{AB}),0]$ and  equality only holds for ideal GKP qudits and lossless transmission ($\eta=1$). Hence, the hashing rate for $\tilde{\rho}_{\boldM_1,\boldM_2}$ would be an achievable distillable entanglement generation rate for the complete noisy state $\rho_{\boldM_1,\boldM_2}$ - we limit our analysis to $I(\tilde{\rho}_{\boldM_1,\boldM_2})$ in the main text as well as the Methods.
}

\subsection{Hashing Rate Asymptote for Low Loss Transmission with Ideal GKP States}
Assuming we start with the state in Eq.~\eqref{eq:ideal_init_state}, where the GKP qudit is an approximate but pure state with variance of $\sigma^2$, the probability of making $k$-shift errors in the logical measurement for sq-GKP states (along either the Pauli $X$ or $Z$ basis) is given by~\cite{Schmidt2024-br},
\begin{align}
         P^{(\mathrm{sq})}(X^{k},\sigma^2)\equiv P^{(\mathrm{sq})}(Z^{k},\sigma^2) &= \sum_{j\in\mathbb{Z}} \frac{1}{2} \left( \mathrm{erf}\left(\sqrt{\frac{2\pi}{d}} \frac{jd+k+1/2}{\sigma}\right) - \mathrm{erf}\left(\sqrt{\frac{2\pi}{d}} \frac{jd+k-1/2}{\sigma}\right)\right) \nonumber ;\\
         &\coloneqq {P}^{(\mathrm{sq})}_{\mathrm{shift}}(k;\sigma^2);  \; k\in \{-\lfloor d/2\rfloor + 1,-\lfloor d/2\rfloor + 2,\ldots, \lfloor d/2\rfloor \},
         \label{eq:shift_prob_methods}
\end{align}
where $k\in\{0,1,2,\ldots,d-1\}$, with $k=0$ denoting the probability of deducing a correct logical outcome, whereas the summation over the index $j$ accounts for shifts of magnitude greater than one lattice spacing. Assuming that Bell state measurement logical outcomes are $l_Z,l_X$ the final state (assuming the twirled noise model) can be expressed as
\begin{align}
	\tilde{\rho}_{\boldsymbol{M}_1 \boldsymbol{M}_2} =  \sum_{k_1,k_2=0}^{d-1} P_{\mathrm{shift}}^{(k_1,k_2)}(\sigma^2)\,  \outprod{\Psi_{k',l'}}_{\boldsymbol{M}_1 \boldsymbol{M}_2} ,
\end{align}
where $P_{\mathrm{shift}}^{(k_1,k_2;)}(\sigma^2)=P(X^{k_1},\sigma^2)\cdot P(Z^{k_2},\sigma^2) $ and the state $\ket{\Psi_{k',l'}}\equiv W_1^{(0,d-l_Z)}W_2^{(d-l_X,0)} \ket{\Psi_{0,0}} $ is a qudit Bell state. The hashing bound~\cite{Lloyd1997-hn,Devetak2005-hi} based lower bound to the distillable entanglement for this state is given by
\begin{align}
	I(\tilde{\rho}_{\boldsymbol{M}_1 \boldsymbol{M}_2}) = \log_2(d) + \sum_{k_1,k_2=0}^{d-1} P_{\mathrm{shift}}^{(k_1,k_2)}(\sigma^2)\log_2 P_{\mathrm{shift}}^{(k_1,k_2)}(\sigma^2).
\end{align}

We now make the assumption that by truncating the summation in Eq.~\eqref{eq:shift_prob_methods} at $j=0$, we can get a sufficient lower bound to the shift error probability, which we label $\tilde{P}(X^k,\sigma^2)$. 
For infinitely squeezed initial states $(\sigma_0^2\rightarrow0)$ being transmiited over a low loss link $(\eta\rightarrow1)$, it is also appropriate to make the assumption that $\tilde{P}_{\mathrm{shift}}({k};\sigma^2)\approx 0$ for $|k|>1$, i.e.\ the probability of incurring larger (greater than one shift) logical errors is negligible. We may now carry out a Taylor expansion of the error function around $d\rightarrow\infty$ (since we expect the system size requirements to increase) and $\eta\rightarrow1$ to yield,
\begin{align}
    	\tilde{P}^{(\mathrm{sq})}_{\mathrm{shift}}(k;1-\sqrt{\eta})\approx	\frac{1}{2} \left( \frac{\left|k+1/2\right|}{k+1/2} - \frac{\left|k-1/2\right|}{k-1/2} + \frac{\sqrt{d\varepsilon} \exp\left(-2\pi(k-1/2)^2 /(d\varepsilon)\right)} {\sqrt{2}\pi(k-1/2)} - \frac{\sqrt{d\varepsilon} \exp\left(-2\pi(k+1/2)^2 /(d\varepsilon)\right) }{\sqrt{2}\pi(k+1/2)} \right),
\end{align}
where $\varepsilon=1-\sqrt{\eta}$. The Taylor series expansion of $\mathrm{erf}(x-\mu/\sigma)$ around $\sigma\rightarrow 0$ and $x\rightarrow\mu$ is
\begin{align}
    \mathrm{erf}\left(\frac{x-\mu}{\sigma}\right) =\frac{\left(\frac{\sigma }{\sqrt{\pi }}+\mathcal{\mathcal{O}}\left(\sigma ^2\right)\right) e^{-\frac{\mu ^2}{\sigma ^2}+\mathcal{O}\left(\sigma ^2\right)}-\mu}{\mu }+x
   \left(\frac{2}{\sqrt{\pi } \sigma }+\mathcal{O}\left(\sigma ^2\right)\right) e^{-\frac{\mu ^2}{\sigma ^2}+\mathcal{O}\left(\sigma ^2\right)}+\mathcal{O}\left(x^2\right),
\end{align}
We then calculate the probabilities of no errors ($k=0$) and one logical error ($|k|=1$) giving us the simplified and approximate shift error expressions,
\begin{align}
    \begin{split}
        &\tilde{P}^{(\mathrm{sq})}_{\mathrm{shift}}(0;\varepsilon)  = 1- \frac{\sqrt{2d\varepsilon} \exp\left(-\pi /(2d\varepsilon)\right)} {\pi} ; \\
     &\tilde{P}^{(\mathrm{sq})}_{\mathrm{shift}}({\pm 1};\varepsilon) =  \frac{\sqrt{d\varepsilon} \exp\left(-\pi /(2d\varepsilon)\right)} {\sqrt{2}\pi}.
    \end{split}
\end{align}

We may make the ansatz that $2d\varepsilon=\xi$ where $\xi\in\mathbb{R}$ is an appropriately chosen constant that captures the required qudit dimensionality requirements for low losses to achieve the best possible rate. Hence, choosing an optimal $\xi$ to obtain the best rate scaling is equivalent to minimizing its difference from the channel capacity, i.e.\ we seek to minimize $Q_2(\sqrt{\eta})- I_{\mathrm{LB}} (\xi) $. This may be performed numerically; choosing $\xi_{\mathrm{opt}}\approx 1.642$ gives us a separation of $\sim1.06$ ebits per channel use between $I_{\mathrm{LB}} (\xi)$ and the repeaterless bound $Q_2(\sqrt{\eta})$. 

Further improvements in the entanglement swapping rate can be achieved by using the hex-GKP encoding. In this case, the logical shift error probability is modified to be  
\begin{align}
         P^{(\mathrm{hex})}(X^{k},\sigma^2)\equiv P^{(\mathrm{hex})}(Z^{k},\sigma^2) &= \sum_{j\in\mathbb{Z}} \frac{1}{2} \left( \mathrm{erf}\left(\sqrt{\frac{4\pi}{\sqrt{3}d}} \frac{jd+k+1/2}{\sigma}\right) - \mathrm{erf}\left(\sqrt{\frac{4\pi}{\sqrt{3}d}} \frac{jd+k-1/2}{\sigma}\right)\right) \nonumber\\
         &\coloneqq {P}^{(\mathrm{hex})}_{\mathrm{shift}}(k;\sigma^2); \; k\in \{-\lfloor d/2\rfloor + 1,-\lfloor d/2\rfloor + 2,\ldots, \lfloor d/2\rfloor \},
         \label{eq:shift_prob2_methods}
\end{align}
to account for the larger separation of $\sim \sqrt{4\pi/\sqrt{3}d}$ between logical states. We may then proceed similarly to the analysis carried out for the sq-GKP qubits - by ignoring greater than single lattice position shift errors, as well as making the appropriate Taylor series expansion, we obtain 
\begin{align}
    \begin{split}
        &\tilde{P}^{(\mathrm{hex})}_{\mathrm{shift}}(0;\varepsilon)  = 1- \frac{\sqrt{\sqrt{3}d\varepsilon} \exp\left(-\pi /(\sqrt{3}d\varepsilon)\right)} {\pi};  \\
     &\tilde{P}^{(\mathrm{hex})}_{\mathrm{shift}}({\pm 1};\varepsilon) =  \frac{\sqrt{\sqrt{3}d\varepsilon} \exp\left(-\pi /(\sqrt{3}d\varepsilon)\right)} {2\pi}.
    \end{split}
\end{align}
In this case, an ansatz of $\sqrt{3}d\varepsilon=\xi'$, allows us to come up with a similar expression for $I_{\mathrm{LB}}^{(\mathrm{hex})}(\xi')$ and to optimize the rate lower bound separation from $Q_2(\sqrt{\eta}$. By choosing $\xi'_{\mathrm{opt}}\approx 1.422$ we obtain a separation of $\sim0.85$ ebits per channel -- this matches the expectation that $\xi_{\mathrm{opt}}/ \xi'_{\mathrm{opt}} = \sqrt{3}/2$, i.e. the system size vs.\ loss scaling is proportional to the improvement in the error probability.
 
\begin{figure}[ht]
		\centering
		\includegraphics[width=0.75\linewidth]{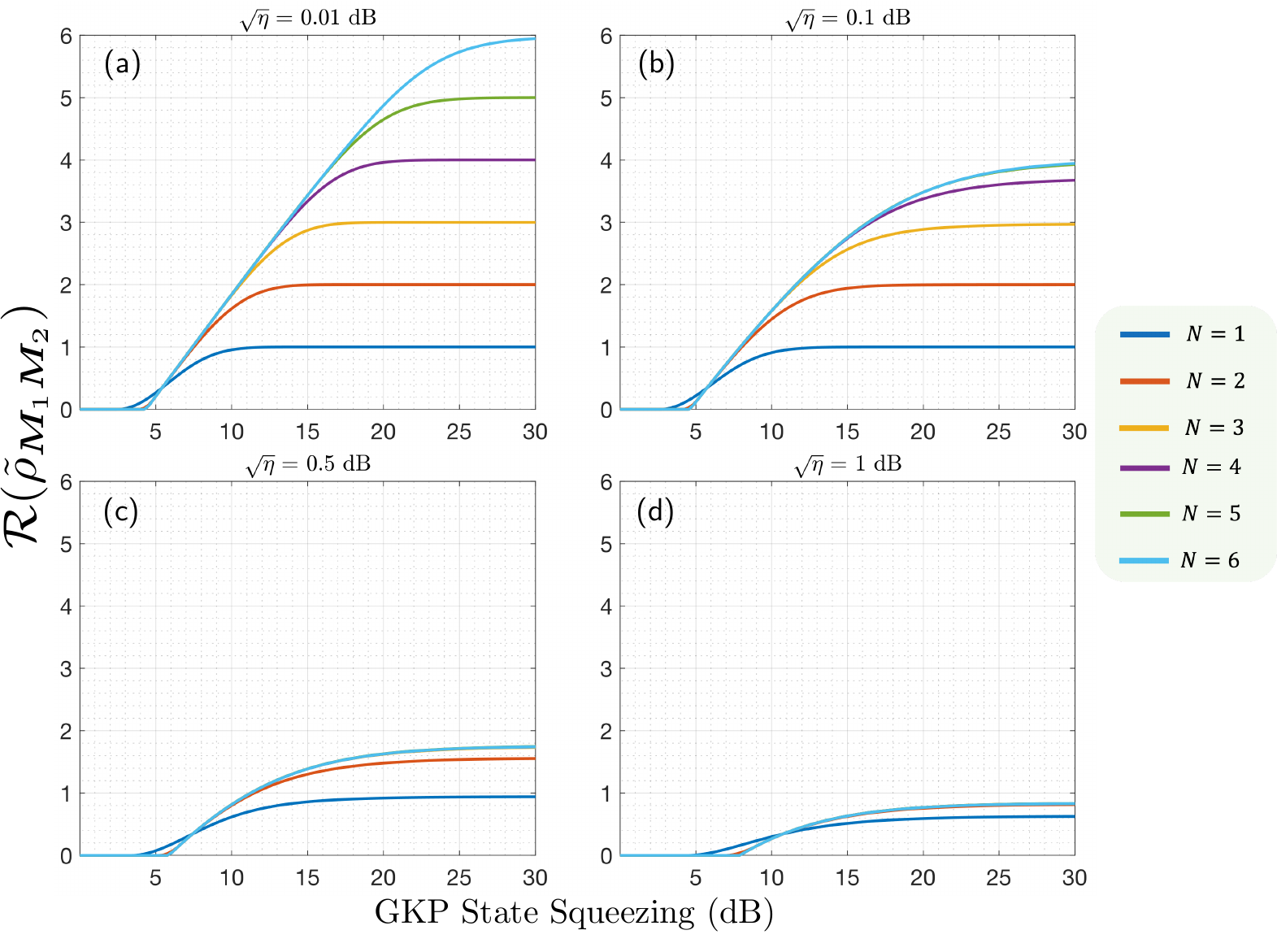}
		\caption{Distillable entanglement rate $\mathcal{R}(\tilde{\rho}_{\boldsymbol{M}_1 \boldsymbol{M}_2})$ for qudit entanglement swap (varying N indicated by line color) at given values of half-channel loss (a) 0.01 dB, (b) 0.1 dB, (c) 0.5 dB and (d) 1 dB
        }
		\label{fig:rate_fixed_loss_v1}
\end{figure}

\begin{figure}[h]
		\centering
		\includegraphics[width=0.75\linewidth]{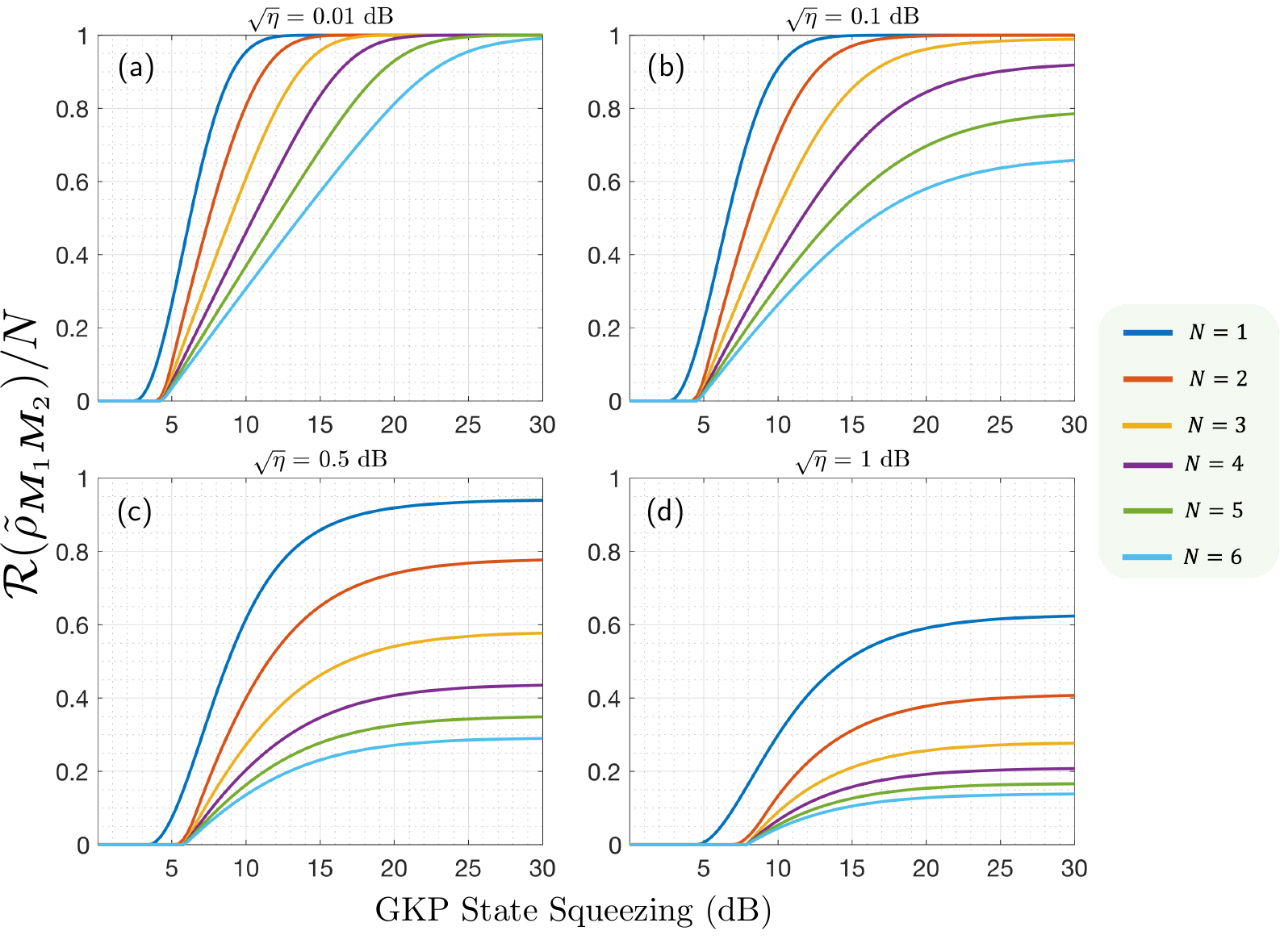}
		\caption{Normalized distillable entanglement rate $\mathcal{R}(\tilde{\rho}_{\boldsymbol{M}_1 \boldsymbol{M}_2})/N $ for qudit entanglement swap (varying $N$ indicated by line color) at given values of half-channel loss (a) 0.01 dB, (b) 0.1 dB, (c) 0.5 dB and (d) 1 dB
        }
		\label{fig:rate_fixed_loss_v2}
\end{figure}

 We may also analyze the hashing rate as a function of the state's squeezing. Figs.~\ref{fig:rate_fixed_loss_v1} and~\ref{fig:rate_fixed_loss_v2} show the achievable distillable entanglement rate (absolute and relative scaling respectively) for the dual-homodyne based swap. For a given value of channel loss (panels a-d), we see that the distillable entanglement rate attains a maximum `steady state' value for higher values of squeezing. The upper limit to this rate decreases with increasing channel loss as seen in Fig.~\ref{fig:rate_fixed_loss_v1}. Additionally the fraction of the maximum achievable $\mathcal{R}(\tilde{\rho}_{\boldsymbol{M}_1 \boldsymbol{M}_2})$ (which for a given qudit encoding is $N$) is dependent on the loss and the encoding size. For higher losses, the maximum `steady state' rate fraction (i.e. $\mathcal{R}(\tilde{\rho}_{\boldsymbol{M}_1 \boldsymbol{M}_2})/N$ ) is lower for larger values of $N$.

\twocolumngrid
\bibliographystyle{apsrev4-1}
\bibliography{biblio_spingkp_letter}
	
\end{document}